\documentclass[aps,pra,showpacs,twocolumn,longbibliography]{revtex4-1}

\usepackage{color}
\usepackage[utf8]{inputenc}
\usepackage[english]{babel}
\usepackage{graphicx}
\usepackage{hyperref}
\usepackage{epstopdf}
\usepackage{amsfonts}
\usepackage{amsmath}
\usepackage{bbold}
\usepackage{color}

\newcommand{\ie}{i.\,e.\ }

\newcommand{\cf}{cf.\ }

\newcommand{\cc}{\text{c.c.}}
\newcommand{\fref}[1]{\text{Fig.}~\ref{#1}}
\newcommand{\eref}[1]{\text{Eq.}~\eqref{#1}}

\begin{document}

\title{Spontaneous crystallization of light and ultracold atoms}
\author{S. Ostermann}
\email{stefan.ostermann@uibk.ac.at}
\author{F. Piazza}
\author{H. Ritsch}
\affiliation{Institut f\"ur Theoretische Physik, Universit\"at Innsbruck, Technikerstraße 21, A-6020 Innsbruck, Austria}

\date{\today}

\begin{abstract}
Coherent scattering of light from ultracold atoms involves an exchange of energy and momentum introducing a wealth of non-linear dynamical phenomena. As a prominent example particles can spontaneously form stationary periodic configurations which simultaneously maximize the light scattering and minimize the atomic potential energy in the emerging optical lattice. Such self-ordering effects resulting in periodic lattices via bimodal symmetry breaking have been experimentally observed with cold gases and Bose-Einstein condensates (BECs) inside an optical resonator.
Here we study a new regime of periodic pattern formation for an atomic BEC in free space, driven by far off-resonant counterpropagating and non-interfering lasers of orthogonal polarization. In contrast to previous works, no spatial light modes are preselected by any boundary conditions and the transition from homogeneous to periodic order amounts to a crystallization of both light and ultracold atoms breaking a continuous translational symmetry. In the crystallized state the BEC acquires a phase similar to a supersolid with an emergent intrinsic length scale whereas the light-field forms an optical lattice allowing phononic excitations via collective back scattering, which are gapped due to the infinte-range interactions.
The studied system constitutes a novel configuration allowing the simulation of synthetic solid state systems with ultracold atoms including long-range phonon dynamics.
\end{abstract}

\maketitle

\section{Introduction}

For a gas of point-like particles off resonantly illuminated by coherent light, the individual dipoles oscillate in phase, each emitting radiation in a characteristic pattern. When several particles contribute to the scattering, the corresponding amplitudes interfere, which leads to a strongly angle-dependent scattering distribution~\cite{andreev1980collective,andreev1980collective, mekhov2007light,zoubi2010metastability}.
In addition, if the motional degree of freedom is relevant on the considered time scales, any high field seeking particle will be drawn towards the corresponding local light field maxima, where in turn light scattering is enhanced. This directional energy and momentum transfer between the gas and the field leads to an instability resulting in density fluctuations and potentially also in the formation of an ordered pattern.  While for a room temperature gas this typically occurs only at very high pump powers~\cite{braun1995self,kartashov2013mid, zheltikov2012nonlinear}, it can become important for very strong scatterers as larger nano- or microparticles~\cite{singer2003self,burns1990optical,tatarkova2002one,demergis2012ultrastrong,karasek2008long,karasek2009longitudinal,dholakia2010colloquium}.
The stringent threshold conditions can be relaxed by laser cooling the gas to temperatures well below the $mK$-range as well as by recycling the scattered light in optical resonators. In this case much weaker forces and thus lower light power is needed to a create substantial back-action effect of the scattered light onto the particles. This back-action was predicted to lead to roton-like instabilities and spatial bunching even at moderate pump powers, as observed in several configurations~\cite{bonifacio1994collective,saffman1998self,inouye1999superradiant,piovella2001superradiant,o2003rotons,kuga_2005,muradyan2005absolute,zimmermann_2007,greenberg2011bunching,schmittberger2016spontaneous,labeyrie2014optomechanical,robb_BEC_2015}. 

A relevant question is thus whether these instabilities can in some cases lead to the formation of a stable crystalline phase in the steady state of such driven, dissipative systems.
The first and simplest instance of such crystals is the self-ordered phase of transversally driven atoms in optical resonators~\cite{domokos2002collective,vuletic_2003,barrett_2012,ritsch2013cold}, with the corresponding transition observable also as a quantum phase transition at zero temperature~\cite{baumann2010dicke,kessler2014steering}. It has been shown recently that a similar phase is also realizable in longitudinally pumped ring-cavities~\cite{ostermann2015atomic}.

While this self-ordered phase shows some aspects shared by standard crystals like a roton-like mode~\cite{mottl2012roton}, other characteristic features like the breaking of a continuous translational symmetry and a crystal spacing which is not externally fixed are both missing, since the resonator mirrors select a single electromagnetic mode. In order to include such features one necessarily needs to couple the particles to several electromagnetic modes, ideally a continuum. This is the case in one-dimensional tapered optical nanofibers~\cite{griesser2013light,chang2013self} or confocal cavities~\cite{gopalakrishnan2009emergent}, where transversally driven atoms are predicted to spontaneously break the continuous symmetry into a crystal phase.
The existence of a continuum of electromagnetic modes opens up the possibility for photons to crystallize, as it was studied with light propagating under Electromagnetically-Induced-Transparency conditions through a nonlinear medium \cite{chang2008crystallization,otterbach_2013}.

\begin{figure}
\centering
\includegraphics[width=0.45\textwidth]{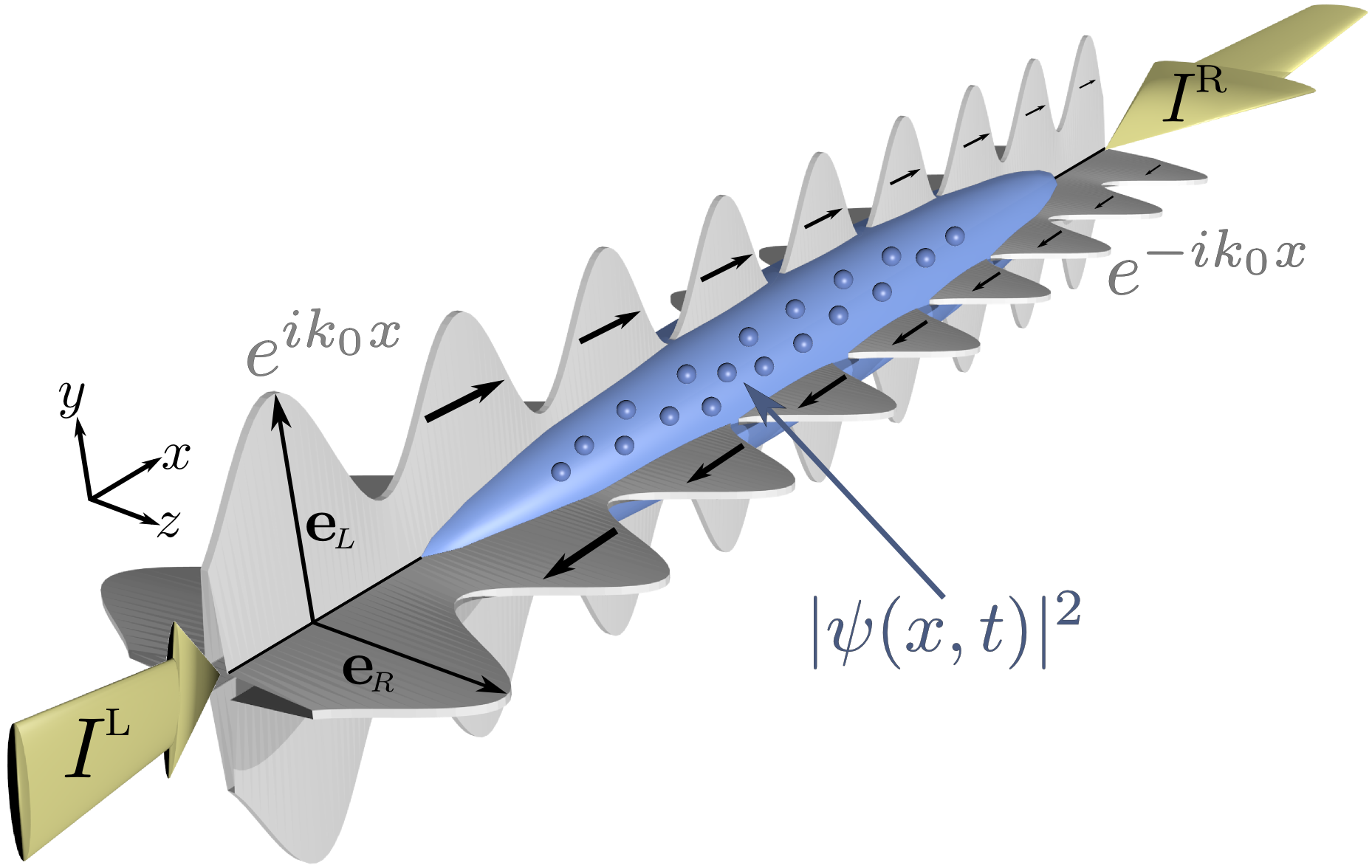}
\caption{Schematic representation of the considered setup. An elongated BEC interacting with two counterpropagating, non-interfering laser beams of orthogonal polarization. The two beams are far detuned from any atomic resonance in order to avoid mixing between the two polarizations. Both polarizations are assumed to be equivalent with respect to the considered atomic transition, the latter thus involving a spherically (or at least cylindrically) symmetric ground-state. Alternatively to the use of two different polarizations, sufficiently different frequencies of the two counterpropagating lasers can be chosen.}
\label{fig:model}
\end{figure}

In this work, we propose and characterise a novel crystalline phase of light and ultracold atoms.
We consider a mirror symmetric and translation invariant setup as it is depicted in~\fref{fig:model}. It involves an elongated Bose-Einstein condensate (BEC) longitudinally illuminated by two counter propagating Gaussian beams far detuned from any atomic resonance. The beams have either orthogonal polarization or a sufficiently large frequency difference to suppress any interference effects. Above a finite driving intensity both atoms and light break a continuous translational symmetry leading to pattern formation with an intrinsically defined lattice spacing determined by the polarizability and density of the gas. The resulting state corresponds to a supersolid BEC trapped in an emerging optical lattice, the latter showing collective phononic excitations. The appearance of an emergent length scale in combination with lattice phonons - i.e. the appearance of a crystal of light - is a crucial difference to configurations where the drive is transverse to the direction in which the system organises~\cite{chang2013self,griesser2013light,gopalakrishnan2009emergent}.

A useful property of the chosen geometry is that ample information about the coupled system dynamics can be retrieved from the reflected light fields in a completely non-invasive manner.  
The present study opens a new direction in (ultra)cold atom-lattice physics, naturally including long range phonon-type interactions and real-time non-destructive monitoring.

\section{Model}\label{sec:model}
We consider a trapped atomic BEC interacting with the electromagnetic (EM) field driven by two far off-resonant, counterpropagating, orthogonally polarized laser beams, as depicted in~\fref{fig:model}. In the dispersive regime considered below, the EM field provides an optical potential for the BEC (see Eq.~(\ref{eqn:GPE})), while the BEC significantly modifies the refractive index (see Eq.~(\ref{eqn:Helmh_total})), thus both field and matter are dynamical quantities.


The BEC is treated within the Gross-Pitaevskii (GP) mean-field approximation \cite{string_pit}, whereby the condensate wave function satisfies the equation
\begin{multline}
i\hbar \frac{\partial}{\partial t}\psi(x,t)=\left[\frac{-\hbar^2}{2m}\frac{\partial^2}{\partial x^2} +V(x)\right]\psi(x,t)\\
+\frac{g_cN}{A}|\psi(x,t)|^2\psi(x,t),
\label{eqn:GPE}
\end{multline}
where $m$ denotes the particle's mass, $g_c$ is the effective s-wave atom-atom interaction strength, and $N$ is the atom number . For computational simplicity we assume the BEC to be confined by an extra transverse trapping potential $V_{\rm trap}(x,y,z)$ such that the dynamics along the $y$ and $z$ axis is negligible. Therefore, the BEC wave function $\psi$ is assumed to be in the ground state of the transverse trap with characteristic size $d_y=d_z=\sqrt{A}$, where  $A$ denotes the BEC cross section. Such a quasi one-dimensional treatment is eligible if the BEC's chemical potential $\mu$ is much smaller than the characteristic transverse trap frequency: $\mu\ll\hbar\omega_{y,z}$. The wavefunction satisfies the normalization condition: $\int dx|\psi(x,t)|^2=1$.

The total optical potential for the BEC has two contributions: 
\begin{equation}
V(x)=V_{\rm trap}(x)+V_{\rm opt}(x),
\label{eqn:Vtot}
\end{equation}
representing the \emph{static} trapping potential $V_{\rm trap}$ and the longitudinal (along x) optical potential $V_{\rm opt}$ determined by the \emph{dynamical} part of the injected and scattered EM field (see Eq.~\eqref{eqn:Vopt}). The latter consists of two far off-resonant fields with orthogonal polarizations driven from the left (L) and right (R) side of the BEC as depicted in~\fref{fig:model}. The two polarization components of the field satisfy the Helmholtz Eq.~\eqref{eqn:Helmh_total}

The atoms inside the BEC are described as linearly polarizable particles with a scalar polarizability $\alpha$ where the imaginary part is negligibly small, \ie spontaneous emission of the atoms is neglected. This corresponds to the assumption that the driving laser frequency $\omega_l$ is sufficiently far detuned form any atomic resonance to prevent substantial internal excitation. This avoids spontaneous emission and thus mixing of the two counterpropagating EM components via Raman scattering as it is used for near resonant polarization gradient cooling may be neglected.

While for spin-polarized atoms the polarizability is field direction dependent in general, we assume the same polarizability for both polarizations orthogonal to the laser axis being the quantization axis. This corresponds to transitions from a spherically (or at least cylindrically) symmetric atomic ground-state. The impinging laser fields from left and right are approximated by plane waves so that we can write the EM field components as $\mathbf{E}_{\rm L,R}(x,t)=\left(E_{\rm L,R}(x)e^{i\omega_l t}+\cc \right)\mathbf{e}_{L,R}$ with the orthogonality condition $\mathbf{e}_L\cdot\mathbf{e}_R=0$. As the light transit time through the sample is negligible compared to all other time scales, the propagation delay of the EM field is adiabatically eliminated and the two field envelops ($\mathrm{L}$ for the field from left and $\mathrm{R}$ for the field from right) satisfy the Helmholtz equations
\begin{align}
\frac{\partial^2}{\partial x^2}E_\mathrm{L,R}(x)+k_0^2 (1+\chi(x))E_\mathrm{L,R}(x)=0
\label{eqn:Helmh_total}
\end{align}
with the wavenumber $k_0$ of the incoming beams and the susceptibility $\chi(x)$ of the BEC. This susceptibility depends on the condensate's density and is given by
\begin{equation}
\chi(x)=\frac{\alpha N}{\epsilon_0A}|\psi(x)|^2,
\label{eqn:susc}
\end{equation}
where $\psi(x,t)$ is the solution of~\eref{eqn:GPE}. The directionality of the field propagation in the Helmholtz equations~\eqref{eqn:Helmh_total} is defined by the boundary conditions, according to which the L-component has a finite imposed amplitude on the left end of the system and the R-component has such on the right end (see also Appendix \ref{app:numerics}). 

As soon as one knows the spatial distribution of the electric fields, one can calculate the optical potential for the atoms via
\begin{equation}
V_\mathrm{opt}(x)=-\frac{\alpha}{A}\left(|E_\mathrm{L}(x)|^2+|E_\mathrm{R}(x)|^2\right).
\label{eqn:Vopt}
\end{equation}
Inserting the optical potential~\eqref{eqn:Vopt} into~\eref{eqn:GPE} leaves us with the set of three coupled differential equations: ~the GP \eref{eqn:GPE} and the two Helmholtz \eref{eqn:Helmh_total}, describing the nonlinear dynamics of our system. 
The degree of nonlinearity resulting from the atom-light coupling is quantified by the dimensionless constant $\zeta$ defined as
\begin{equation}
\zeta:=\frac{\alpha N}{\varepsilon_0\lambda_0 A}=\frac{\alpha}{\epsilon_0}n\frac{L}{\lambda_0}\;,
\end{equation}
where $n=N/AL$ is the three-dimensional density of the homogeneous BEC with $L$ its characteristic extension along $x$.
Due to the adiabatic approximation involved in the Helmholtz equation, the EM fields depend only parametrically on time through the dynamical refractive index set by the BEC density.

Due to the orthogonality of the two chosen polarizations there is no interference between the two counterpropagating components of the EM fields. Therefore, the optical potential~\eqref{eqn:Vopt} only depends on the absolute value squared of the fields. This important feature guarantees the translation invariance of the setup along the x direction nevertheless maintaining a mirror symmetric setup. Indeed, since we are driving with plane-wave lasers, as long as the BEC density is homogeneous, the EM fields $E_{\rm L,R}(x)$ in Eq.~(\ref{eqn:Helmh_total}) are also plane waves, leading to a translation invariant optical potential~\eqref{eqn:Vopt}. This invariance with respect to continuous translations is spontaneously broken above a finite driving intensity, as discussed in section~\ref{sec:crys_thres}. In the resulting crystalline phase, the lattice constant is intrinsically established as it is discussed in section~\ref{sec:groundstate}. This is due to the fact that no specific modes are selected and the fields can counterpropagate independently.

\section{Dynamical instability towards crystallisation}
\label{sec:crys_thres}

As already mentioned above, due to the orthogonality of the polarizations of the two injected counter-propagating laser fields the particles do not feel any longitudinal optical forces.  Naively, one could thus expect the BEC to remain unperturbed independently of the pump intensity. In this section we show that this is actually not the case, as above a particular threshold driving strength small density fluctuations lead to backscattering of light which in turn amplifies these fluctuations. This leads to an instability towards crystallization in the longitudinal direction. The latter can be described by considering the collective excitation spectrum of the system for a spatially homogeneous density distribution of the BEC $\psi_0(x,t)=1/\sqrt{L}$ with the corresponding propagating field solution of Eq.~\eqref{eqn:Helmh_total}. These are plane waves of the form $E_{\rm L,R}^{(0)}=C \exp(\pm i k_{\rm eff}x)$, with the modified wavenumber
\begin{equation}
\label{eqn:keff}
k_{\rm eff}=\frac{2\pi}{\lambda_0}\sqrt{1+\zeta\lambda_0|\psi_0|^2}=\frac{2\pi}{\lambda_0}\sqrt{1+\frac{\alpha}{\epsilon_0}n},
\end{equation}
where $C$ is a real number fixed by the driving strength.

The spectrum is obtained by linearizing the coupled equations \eqref{eqn:GPE},\eqref{eqn:Helmh_total} with the ansatz $\psi=(\psi_0+\delta\psi)e^{-i\mu t}$ and $E_{\mathrm{L},\mathrm{R}}=E_{L,R}^{(0)}+\delta E_{L,R}$. Here $\delta\psi$ and $\delta E$ are small deviations from the stationary solutions $\psi_0$ or $E_{L,R}^{(0)}$ and $\mu$ is the BEC chemical potential (we refer to Appendix~\ref{app:linearize} for any details).
This yields
\begin{multline}
\hbar^2\omega_q^2=\frac{\hbar^2 q^2}{2m}\bigg[\frac{\hbar^2 q^2}{2m}+2gn\\
-\frac{64\pi^2A\zeta^2}{c N L}\frac{1}{q^2-4 k_{\rm eff}^2}I^{L,R}\bigg].
\label{eqn:exspectr}
\end{multline}
Here $I^{L,R}$ denotes the intensity (in W/m$^2$) of the incoming light which we have chosen to be equal from left and right.

The above analytical expression~\eref{eqn:exspectr} is very useful to understand some essential features of the atom-light interaction in the present setup and in particular the nature of the crystallisation transition. Apart from the last term, we recognize the known form of the Bogoliubov spectrum of interacting BECs~\cite{string_pit}, with the linear-in-$q$ behavior corresponding to phononic excitations at low $q$.
The last term on the other hand is the only one resulting from the atom-light interactions.
The first thing to note is that its denominator vanishes at $q=\pm 2k_{\rm eff}$, which tells immediately that the modified wavenumber~\eqref{eqn:keff} sets the favoured momentum for the appearance of the instability. However, the vanishing of the denominator is compensated by the diverging BEC length $L$, at every finite atom number $N$ (note that $\zeta\sim N$). The limit $L\to\infty$ of~\eref{eqn:exspectr} actually \emph{has} to be taken, since the stationary plane-wave solution $E_{\rm L,R}^{(0)}\sim \exp(\pm i k_{\rm eff}x)$ about which we linearized only makes sense for a homogeneous \emph{and} infinite atomic medium, so that the edges may be neglected. This indeed allows us to neglect the reflection of the incident wave by the change in refractive index at the BEC edges. Such finite-size effects, included in the numerical solutions described in section~\ref{sec:groundstate}, become irrelevant for large systems, as we demonstrate below.
\begin{figure}
\centering
\includegraphics[width=0.4\textwidth]{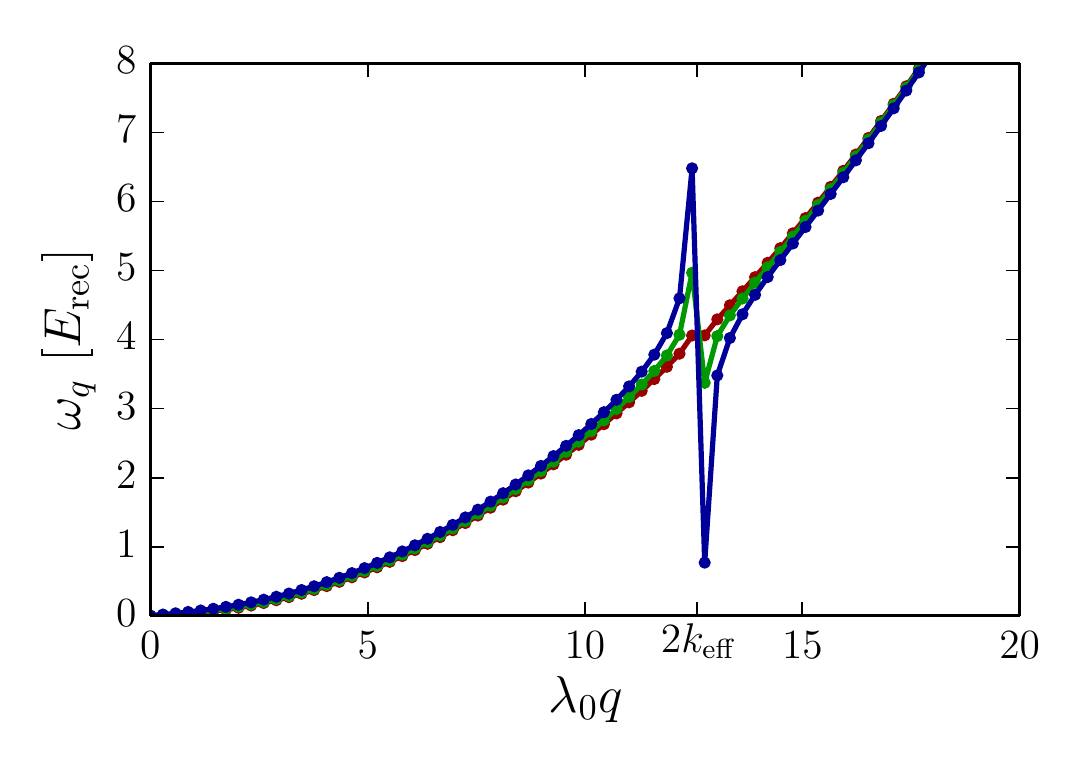}
\caption{Excitation spectrum \eqref{eqn:exspectr} in the homogeneous phase for different field intensities $I^{L,R}=2.0$ (red), $I^{L,R}=20.0$ (green) and $I^{L,R}=60.0$ (blue). ($\zeta=0.1$, $L=100 \lambda_0$, $g_c N/A\lambda_0=E_{\rm rec}$)}
\label{fig:spectrum}
\end{figure}

One way to obtain the proper result for~\eref{eqn:exspectr} in the limit $L\to\infty$ is to consider that for any finite $L$ the allowed momenta $q$ take only quantized values as multiples of $2\pi/L$. Before taking the limit $L\to\infty$ it is instructive to compute the spectrum~\eqref{eqn:exspectr} for fixed finite $L$, as it is shown in~\fref{fig:spectrum} for different values of  $I^{L,R}$. One recognizes a gap opening at $q=2 k_{\rm eff}$ for any finite $I^{L,R}$, i.e. any finite driving strength. The spectrum develops a minimum at the finite momentum $q=2 k_{\rm eff}+(2\pi)/L$, which corresponds to a roton minimum in the language commonly adopted for standard crystal formation~\cite{chaikin_lub}. It constitutes a generalization to continuous-symmetry breaking of the roton-like instability observed with a BEC in an optical cavity~\cite{eth_soft}.
In a similar manner as in standard crystals, the crystallisation threshold can be calculated by finding the drive intensity at which the roton energy approaches zero. This leads to the threshold-condition $\omega_{2k_\mathrm{eff}+(2\pi)/L}=0$.
\begin{figure}
\centering
\includegraphics[width=0.35\textwidth]{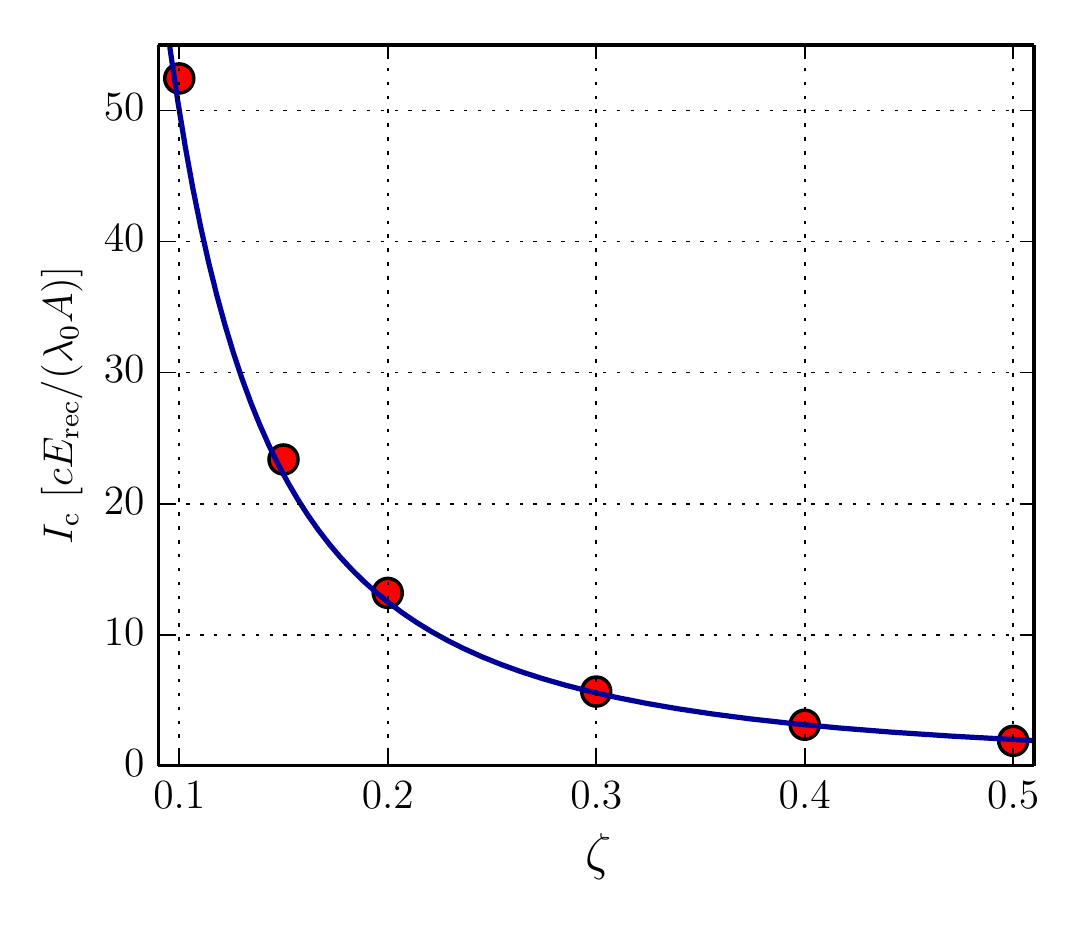}
\caption{$\zeta$ dependence of the critical intensity. The solid blue line depicts the analytical result defined by~\eref{eqn:Icrit} whereas the red dots depict numerical threshold estimations for large system sizes ($L=120\lambda_0$). ($\zeta=0.1$, $g_c N/A\lambda_0=E_{\rm rec}$)}
\label{fig:threshold_zeta_dep}
\end{figure}
We are now in the position to take the limit $L\to\infty$. In doing this we note that we have to keep the atom number $N$ constant in order to get a finite critical drive strength. Otherwise, if we perform the standard thermodynamic limit $N/L=\text{const.}$, the energy of the system diverges and the crystallisation threshold vanishes. This divergence is an artefact of our model in which the light-mediated atom-atom interaction is of infinite range since the EM field is adiabatically adapting to the BEC configuration. The inclusion of the dynamics of the EM field (retardation effects) would introduce a finite range and thus eliminate the divergence in the energy. Still, the resulting range is expected to be larger than the typical BEC size $L$ so than our calculation should be valid for any realistic system size.
Taking the $L\to\infty$ limit we thus get the critical driving intensity
\begin{equation}
I_{c}^{\rm L,R}=\frac{cE_{rec}N}{\lambda_0 A}\frac{1}{\zeta^2}=cE_{rec}\frac{\varepsilon_0^2}{\alpha^2 }\frac{1}{n}\frac{\lambda_0}{L},
\label{eqn:Icrit}
\end{equation}
where we introduced the recoil energy $E_\mathrm{rec}:=\hbar\omega_{rec}=\hbar^2 k_0^2/(2m)$.

Note that in the $L\to\infty$ limit with constant $N$ the BEC becomes more and more dilute, which renders the direct atom-atom coupling $\sim g_c$ eventually irrelevant. 
In~\fref{fig:threshold_zeta_dep} the analytical expression~\eqref{eqn:Icrit} is compared with numerically estimated thresholds for large system sizes (see section~\ref{sec:groundstate}). We find full agreement between the linear instability threshold and the numerical threshold found by studying the imaginary time evolution of eqns.~\eqref{eqn:GPE} and \eqref{eqn:Helmh_total}. This numerical approach to finite-sized systems is described next.

\section{Crystal of light and atoms} \label{sec:groundstate}

After showing that the homogeneous system is unstable above a certain driving intensity, we are going to show that a stable crystalline phase is reached and study its properties by numerically solving the coupled GP~\eqref{eqn:GPE} and Helmholtz~\eqref{eqn:Helmh_total} equations. We perform an imaginary time evolution of the system~\eqref{eqn:GPE}-~\eqref{eqn:Helmh_total},~\ie replace $t\rightarrow i\tau$, which yields the ground state of the system for long enough evolution times.
For a detailed description of the numerical methods we refer to Appendix~\ref{app:numerics}.
\begin{figure}
\centering
\includegraphics[width=0.5\textwidth]{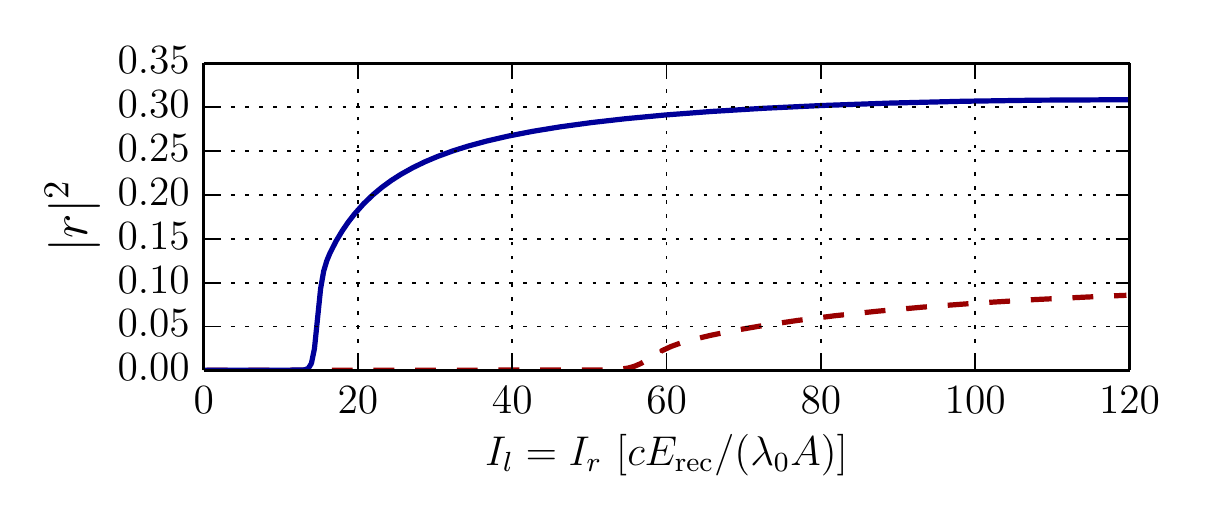}
\vspace{-0.5cm}
\caption{Dependence of the reflection coefficient of the BEC on the incoming field amplitudes for different atom-field couplings $\zeta=0.1$ (dashed red) and $\zeta=0.2$ (solid blue). The remaining parameters are the same as in Fig.~\ref{fig:spectrum}}
\label{fig:threshold}
\end{figure}

To determine the crystal transition point as a function of driving intensity we compute the total reflectivity of the BEC with respect to the intensity of either one of the incident beams, which we again take to be equal. For large enough system sizes, a clear threshold behavior is visible at a critical driving intensity, whereby the reflectivity grows from essentially zero with almost infinite slope,~\cf~\fref{fig:threshold}. The hereby found critical intensity is in perfect agreement with the analytical result(see~\fref{fig:threshold_zeta_dep}).
As mentioned already in the previous section, finite-size effects manifest due to the presence of the edges of the BEC. In the calculations described in this section and in section~\ref{sec:phonons}, there is no further trapping potential along $x$ and the BEC is confined within a box of size $L$, so that the BEC has sharp edges for the light impinging at $x=0$ and $x=L$ (see Appendix~\ref{app:numerics} for more details). In section~\ref{sec:exp} we add an harmonic trap along $x$ and show that the qualitative behavior is the same as described here. The BEC edges create a quick increase of the refractive index which induces a small amount of reflection of the incoming beam. As apparent from~\fref{fig:threshold} this reflection is irrelevant for large system sizes compared to the reflection present in the crystalline phase.

\begin{figure}
\centering
(a) \includegraphics[width=0.45\textwidth]{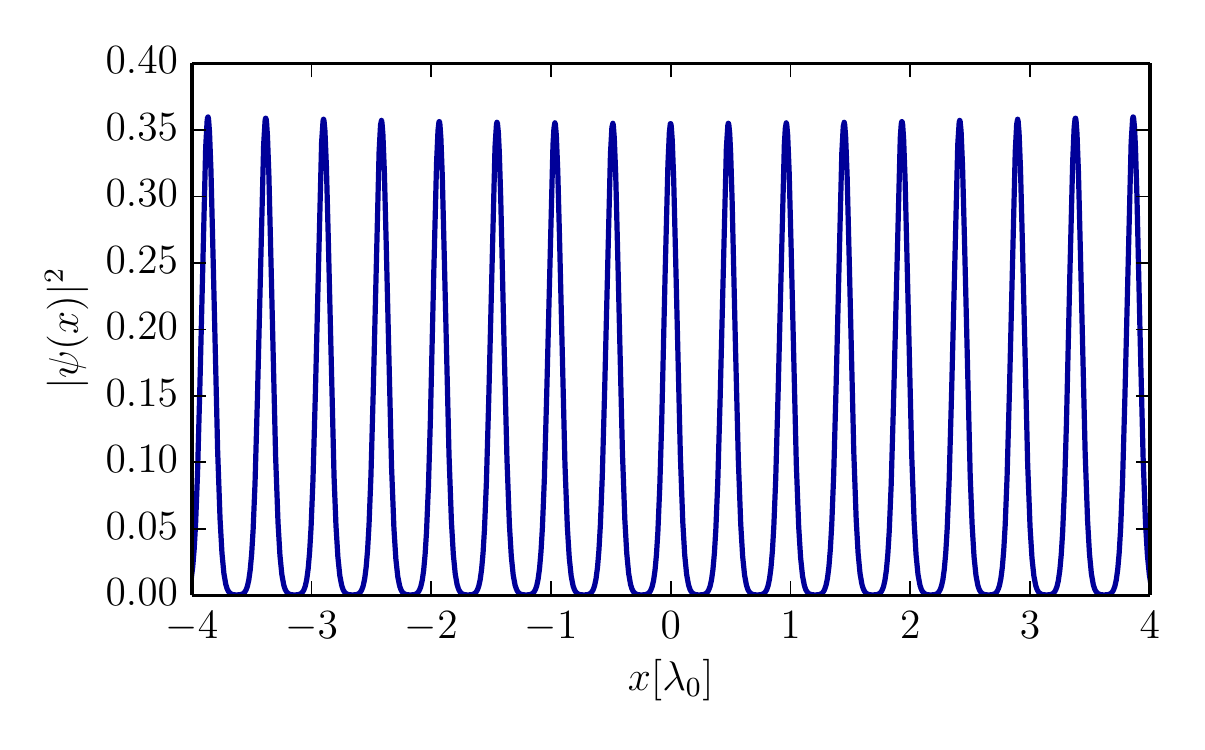}
\\
(b) \includegraphics[width=0.45\textwidth]{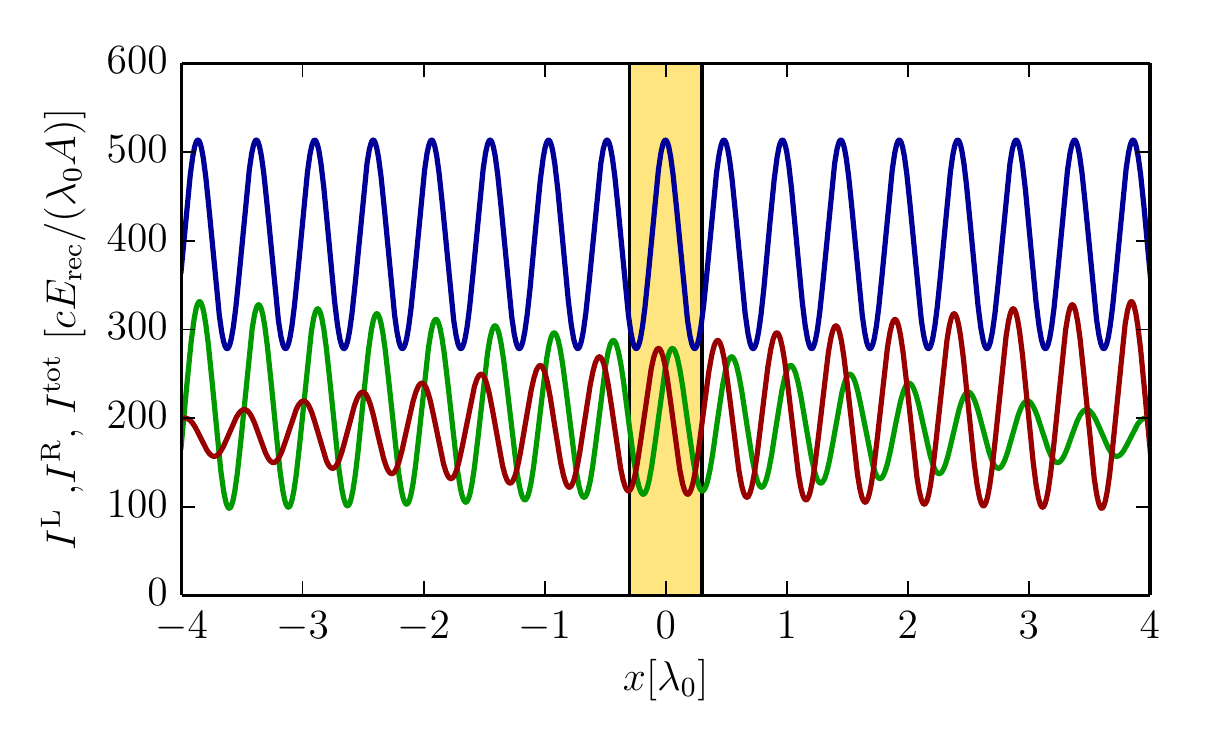}
\caption{(a) Crystal ground state for $\zeta=0.1$, $I_l=I_r=200$ and (b) corresponding intensity distribution for the field from left (green) and right (red). The solid blue line depicts the sum of both intensities. A zoom into the yellow shaded region can be found in~\fref{fig:zoom}. The remaining parameters are $g_c N/A\lambda_0=E_{\rm rec}$ and $L=10\lambda_0$.}
\label{fig:groundstate}
\end{figure}

The large light reflection above threshold is due to the appearance of a large spatial modulation of the BEC, forming the density grating shown in~\fref{fig:groundstate}(a). This corresponds to a continuous symmetry breaking at the threshold leading to a crystalline phase which for the phase-coherent BEC implies supersolid order. Each peak in the density grating reflects the incoming light, resulting in a damped modulation of the intensity of each polarization component across the condensate, as shown in~\fref{fig:groundstate}(b). While the modulation of each component's intensity $I^{\rm L,R}$ is damped across the system, the modulation of the total intensity $I^\mathrm{tot}=I^{\rm L}+I^{\rm R}$ is not damped, resulting in a periodic optical-lattice potential for the BEC which matches its density grating. 

An important feature of the optical lattice emerging in the crystalline phase is the intrinsic character of the lattice spacing, which is not fixed externally but rather set by the BEC density and atom polarizability. This is a clear difference with respect to the self-ordering in optical resonators where the spacing is externally fixed by the cavity mirrors~\cite{domokos2002collective}; and also to the case of self-ordering of transversally driven atoms coupled to the continuum of modes of optical fibers, where the spacing is fixed by the driving frequency and fiber dispersion \cite{chang2013self,griesser2013light}.
As anticipated in section~\ref{sec:crys_thres}, the appearance of the roton-like instability at the characteristic momentum $2k_{\rm eff}$ leads to the following prediction for the emergent lattice spacing:
\begin{align}
\label{eq:spacing}
d=\frac{\pi}{k_{\rm eff}}=\frac{\lambda_0}{2\sqrt{1+\frac{\zeta\lambda_0}{L}}}=\frac{\lambda_0}{2\sqrt{1+\frac{\alpha}{\epsilon_0}n}}.
\end{align}
The emergent spacing is always smaller than the one in vacuum $\lambda_0/2$.
This feature can be qualitatively reproduced also within a toy-model where the medium is approximated by a set of beam splitters~\cite{ostermann2014scattering}.
\begin{figure*}
\centering
\includegraphics[width=0.8\textwidth]{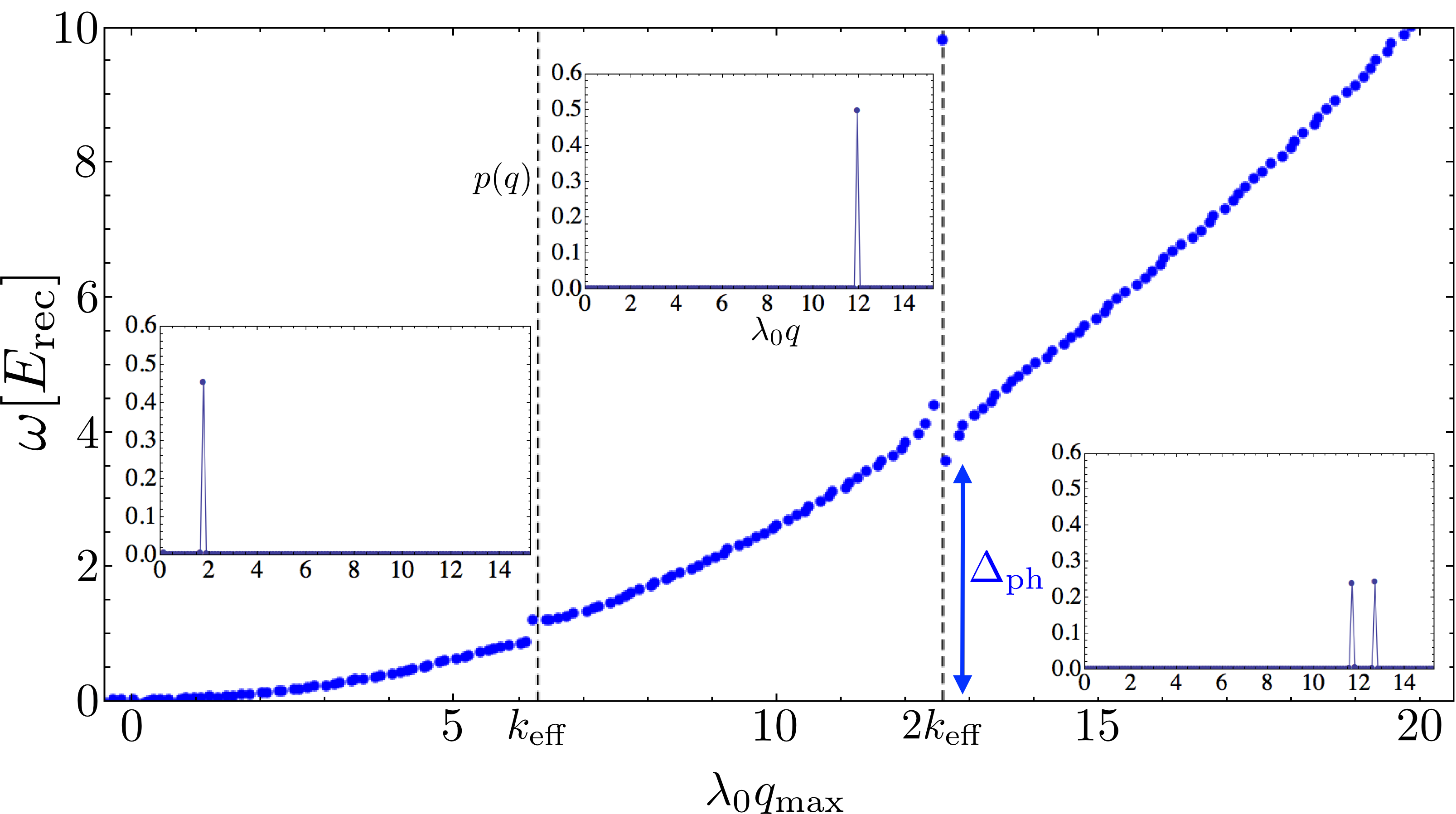}
\caption{Excitation spectrum of the atom-light crystal. The blue points are the eigenvalues of the GP and Helmholtz equations linearized about the crystalline stationary state (see Appendix \ref{app:linearize}). The numerical diagonalization is performed with a momentum-space discretization $dq=2\pi/L$. The parameters are the same as in Fig.~\ref{fig:spectrum} except for $L=50$ and a fixed drive intensity $I^{L,R}=50$ (slightly above threshold). $q_{\rm max}$ is the momentum corresponding to the largest component of the eigenvector of each eigenvalue. The insets show examples of eigenvectors (unormalized probability in momentum space) for three different eigenvalues representative of each region of the spectrum, from left to right: $\lambda_0q_{\rm max}=1.38<\lambda_0k_{\rm eff}$, $\lambda_0k_{\rm eff}<\lambda_0q_{\rm max}=11.9<2\lambda_0k_{\rm eff}$, and $\lambda_0q_{\rm max}=12.7>2\lambda_0k_{\rm eff}$. The latter region corresponds to lattice phonons, characterized by a two symmetric pairs of peaks about a finite momentum. This phononic branch: $q_{\rm max}>2k_{\rm eff}$ has a gap $\Delta_{\rm ph}$. Its analytical estimate in Eq.~(\ref{eq:ph_gap}) yelds $\Delta_{\rm ph}\simeq 2\sqrt{2} E_{rec}$, in reasonable agreement with the numerical data.}
\label{fig:spect_above}
\end{figure*}
This typically small but nonetheless crucial effect is also present when using counterpropagating beams with equal polarization and is essential for atom trapping in optical lattices~\cite{deutsch1995photonic}. Would the atoms indeed be trapped with the vacuum spacing $\lambda_0/2$, the EM field would be perfectly reflected and no standing wave could actually be formed and thus no trapping be possible. It is only through the slight renormalization  $d<\lambda_0/2$ that perfect reflection is avoided. What our scheme with orthogonally polarized counterpropagating beams allows is to make the small renormalization of $d$ coincide with the appearance of a large density modulation out of a homogeneous phase,~\ie a crystallisation.

The existence of an intrinsic lattice spacing in the crystalline phase implies as well the presence of phononic excitations of the the lattice, as discussed in the next section.

\section{Excitations of the crystal: phonons}\label{sec:crystal_excitations}

Further insight in the properties of the atom-light crystal is provided by analyzing its excitation spectrum. As done in Section~\ref{sec:crys_thres}, we linearize the coupled system of Eqs.~(\ref{eqn:GPE}),(\ref{eqn:Helmh_total}). However, now the perturbation is performed around the symmetry-broken stationary solution. The result is presented in Fig.~\ref{fig:spect_above} for a driving intensity slightly above threshold. 
Details of the calculation are given in the Appendix \ref{app:linearize}. Since translation-invariance is
broken, the matrices describing the linear system are not diagonal in momentum space requiring a discretization of the position(momentum) continuum. Moreover, while the total light intensity and atom density are periodic, the intensity of each polarization component is not, due to accumulated reflection along the density grating, introducing the decaying evelope shown in Fig.~\ref{fig:groundstate}(b).  This prevents the use of the quasi-momentum to label the excitation modes. 

In Fig.~\ref{fig:spect_above} we labelled the eigenvalues based on their dominant momentum component $q_{\rm max}$, extracted from the corresponding eigenvector. This allows to split the spectrum into three regions separated by gaps at $q_{\rm max}=k_{\rm eff}$ and $q_{\rm max}=2k_{\rm eff}$. 

The gap at $q_{\rm max}=k_{\rm eff}$ opens up for $I^{L,R}> I_c^{L,R}$ due to the appearance of an optical lattice potential for the atoms with a $\pi/k_{\rm eff}$ periodicity. It separates the two bands which, slightly above threshold, are characterized by eigenvectors with a clearly dominant momentum component (see left and middle inset in Fig.~\ref{fig:spect_above}).

On the other hand, the gap at $2k_{\rm eff}$ is the same one appearing in the homogeneous phase (see Fig.~\ref{fig:spectrum}). As discussed in Section~\ref{sec:crys_thres}, at the critical drive intensity $I_c^{L,R}$ the gap is such that the energy of the mode with momentum $q=2k_{\rm eff}+2\pi/L$ (momentum is still a good quantum number for $I^{L,R}\leq I_c^{L,R}$) vanishes. Out of this zero-energy mode at $2k_{\rm eff}$ (not resolved with the discretization of Fig.~\ref{fig:spect_above}), and beyond the critical point: $I^{L,R}> I_c^{L,R}$, the lattice phonon branch develops for $q_{\rm max}>2k_{\rm eff}$. The momentum distribution of the lattice-phonon eigenvectors is characterized by the splitting of the single peak at $2k_{\rm eff}$ into two neighbouring peaks (see rightmost inset of Fig.~\ref{fig:spect_above}). The phonon wavelength is set by the distance between the two nearby maxima appearing in the momentum distribution. This generates the slow beating in coordinate space. With a finite system size $L$, the longest wavelength is of the order of $L$.

Moreover, the lattice-phonon branch is gapped, in the sense that its lowest energy mode at $q_{\rm max}$ slightly above $2k_{\rm eff}$ has a finite energy, as visible in Fig.~\ref{fig:spect_above}. More importantly, this gap remains finite in the thermodynamic limit $L\to\infty$. We can estimate the size of the lattice-phonon gap close to threshold by using Eq.~(\ref{eqn:exspectr}) and computing the energy of the mode next to the zero-energy mode. This yelds
\begin{align}
\label{eq:ph_gap}
\Delta_{\rm ph}^2\simeq 4\frac{\hbar^2 k_{\rm eff}^2}{2m}\left(2\frac{\hbar^2 k_{\rm eff}^2}{2m}+gn\right)\;,
\end{align}
which takes the value $\Delta_{\rm ph}^2\simeq 8E_{\rm rec}^2$ in the thermodynamic limit $L\to\infty$ with $N=\text{const.}$. As discussed in section~\ref{sec:crys_thres}, in this limit $I_c^{L,R}$ remains finite while $n\to 0$ and $k_{\rm eff}\to k_0$. Another choice of thermodynamic limit is possible: $L,N\to\infty$ with $n=\text{const.}$, where $I_c^{L,R}\to 0$ and the gap is still given by (\ref{eq:ph_gap}). 
The existence of an energy gap for lattice phonons is due to the long-range nature of the interactions, as it can be already predicted within a classical model of interacting point-like particles~\cite{ashcroft1976solid}. 
From a more general field-theoretical perspective, some of the gapless Goldstone modes expected from the continuous-symmetry breaking can indeed disappear (i.e. become gapped) due to the long range of the interactions, as it for instance happens to the longitudinal phonons of a three-dimensional Wigner crystal \cite{wigner1938effects}.
As long as retardation effects can be neglected our interactions will be infinite-ranged, the lattice-phonons gapped, and thus quantum/thermal fluctuations will not destroy crystalline order even in truly one dimension \cite{yukalov_2015}.

The existence of lattice phonons among the collective excitations is confirmed by numerical simulations of the real-time dynamics of the system, as it is described in the next section.

\section{Crystallisation dynamics after a quench}\label{sec:phonons}

\begin{figure}
\centering
(a) \includegraphics[width=0.4\textwidth]{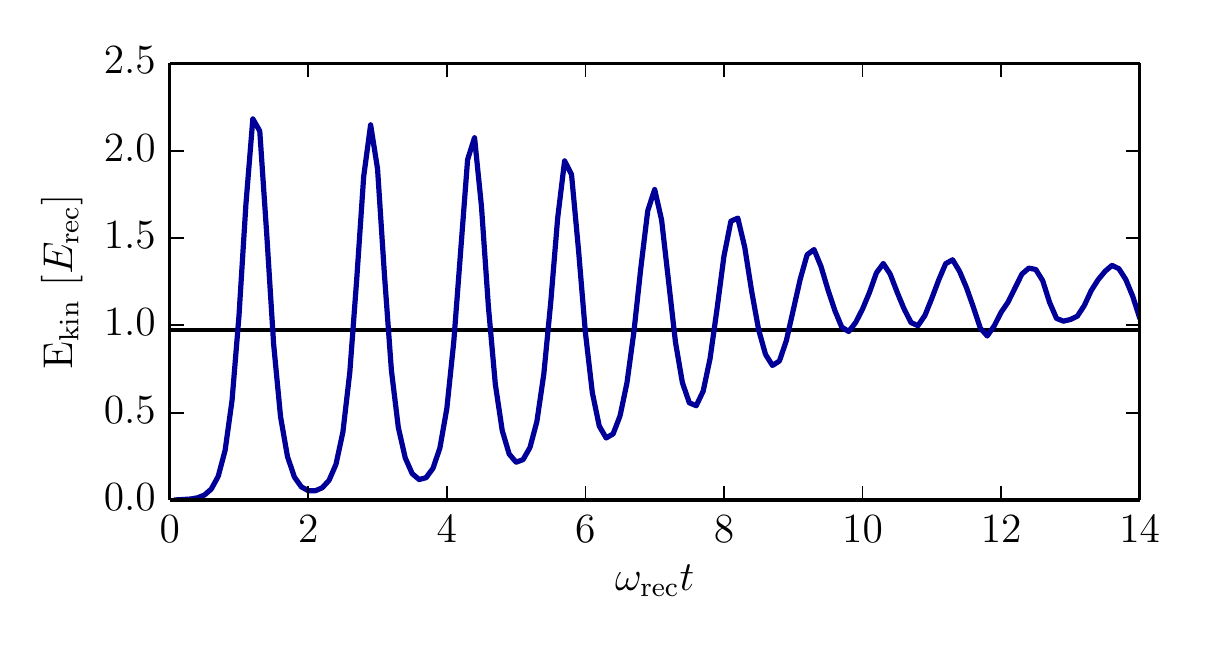}\\
(b) \includegraphics[width=0.4\textwidth]{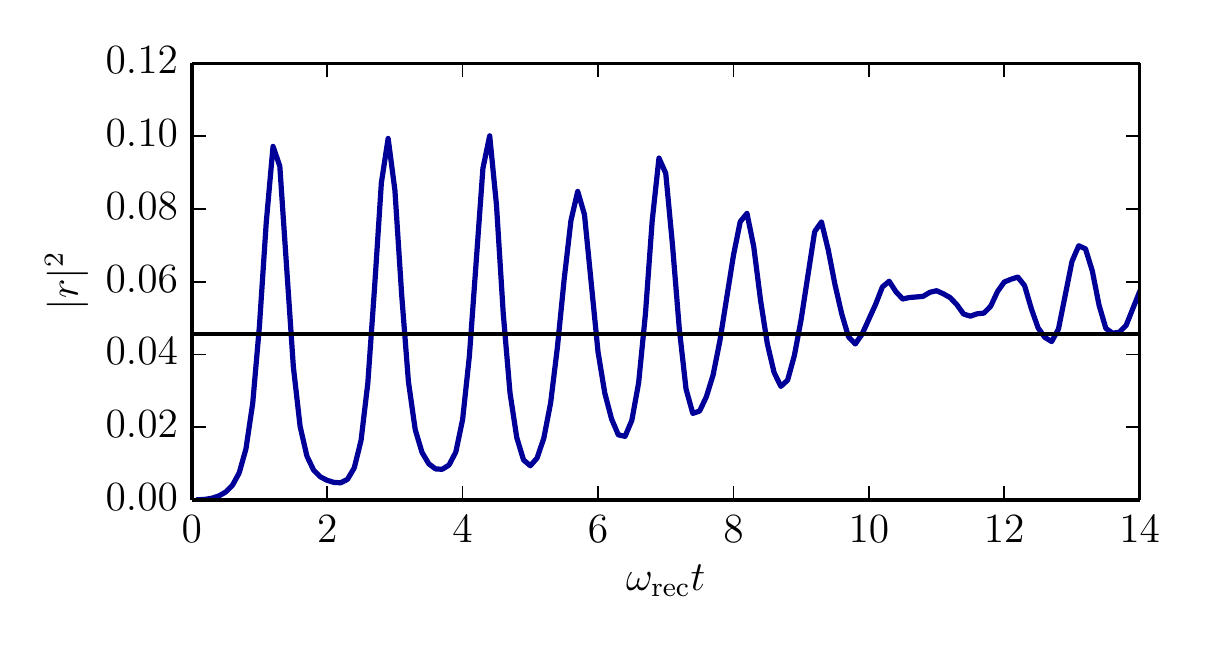}
\caption{(a) Real-time evolution of the kinetic energy for $\zeta=0.1$, $I_l=I_r=100$, $g_cN=1$. (b) Real-time evolution of the reflection coefficient for the same parameters as in figure (a). The solid black line shows the mean value of the corresponding functions.}
\label{fig:Ekin_tdep}
\end{figure}

In this section we investigate the real time dynamics of the system by directly solving eqns.~\eqref{eqn:GPE} and~\eqref{eqn:Helmh_total}. 
This allows us to analyse the crystallization dynamics after a sudden turn on (quench) of the pump laser strength from zero to a value above threshold at $t=0$. The corresponding time evolution of the BEC reflectivity, kinetic energy, as well as the evolution of the BEC density and total light intensity are shown in Figs.~\ref{fig:Ekin_tdep} and~\ref{fig:densint_realt}.

\begin{figure}
\centering
\includegraphics[width=0.4\textwidth]{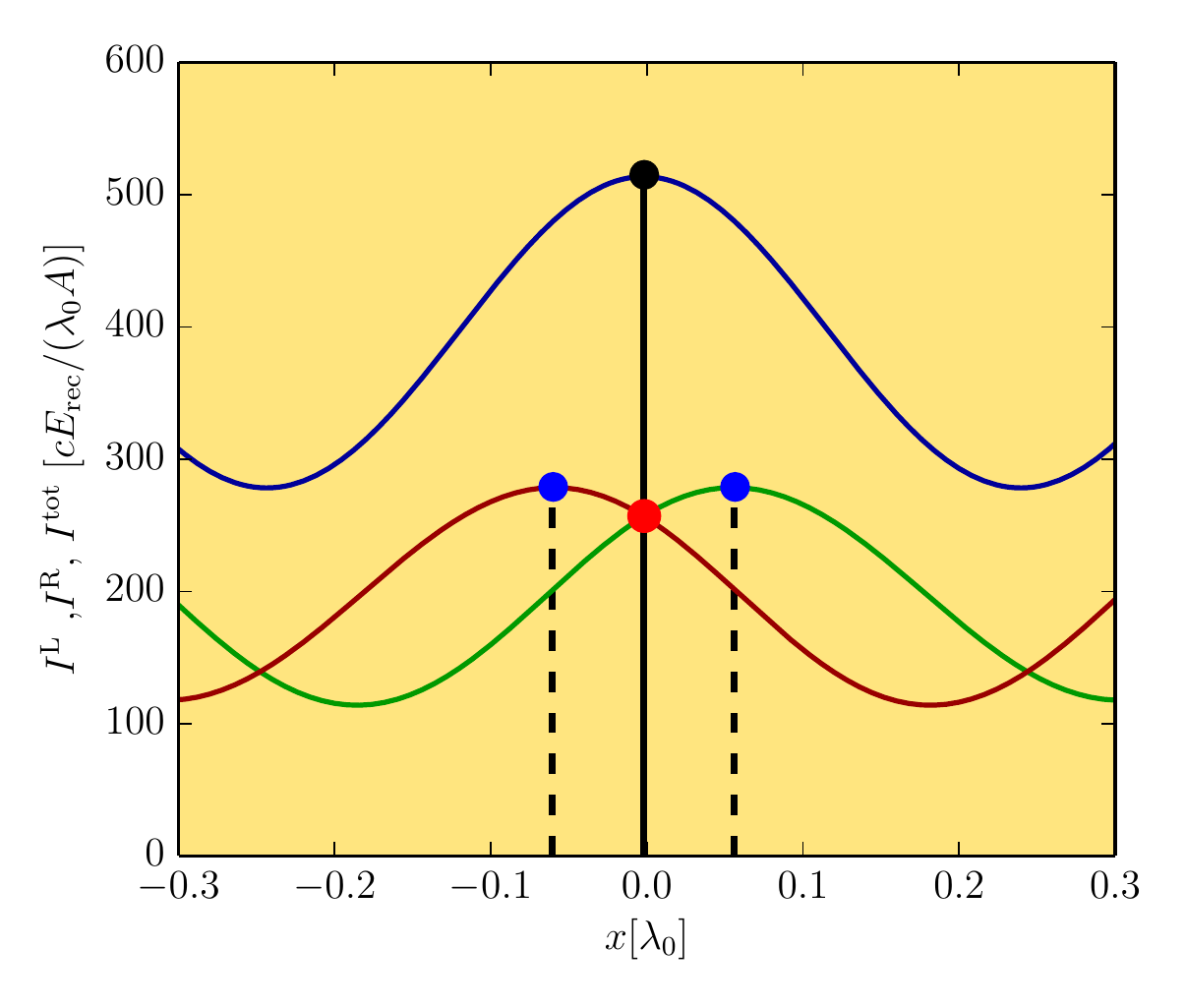}
\caption{Zoom into the yellow shaded region of~\fref{fig:groundstate}. The blue dots mark the maxima of the field from left (green) and right (red) whereas the black dot marks the maximum of the total field intensity (blue). The red dot shows the actual position of the particles.}
\label{fig:zoom}
\end{figure}

\begin{figure}
\centering
(a) \includegraphics[width=0.45\textwidth]{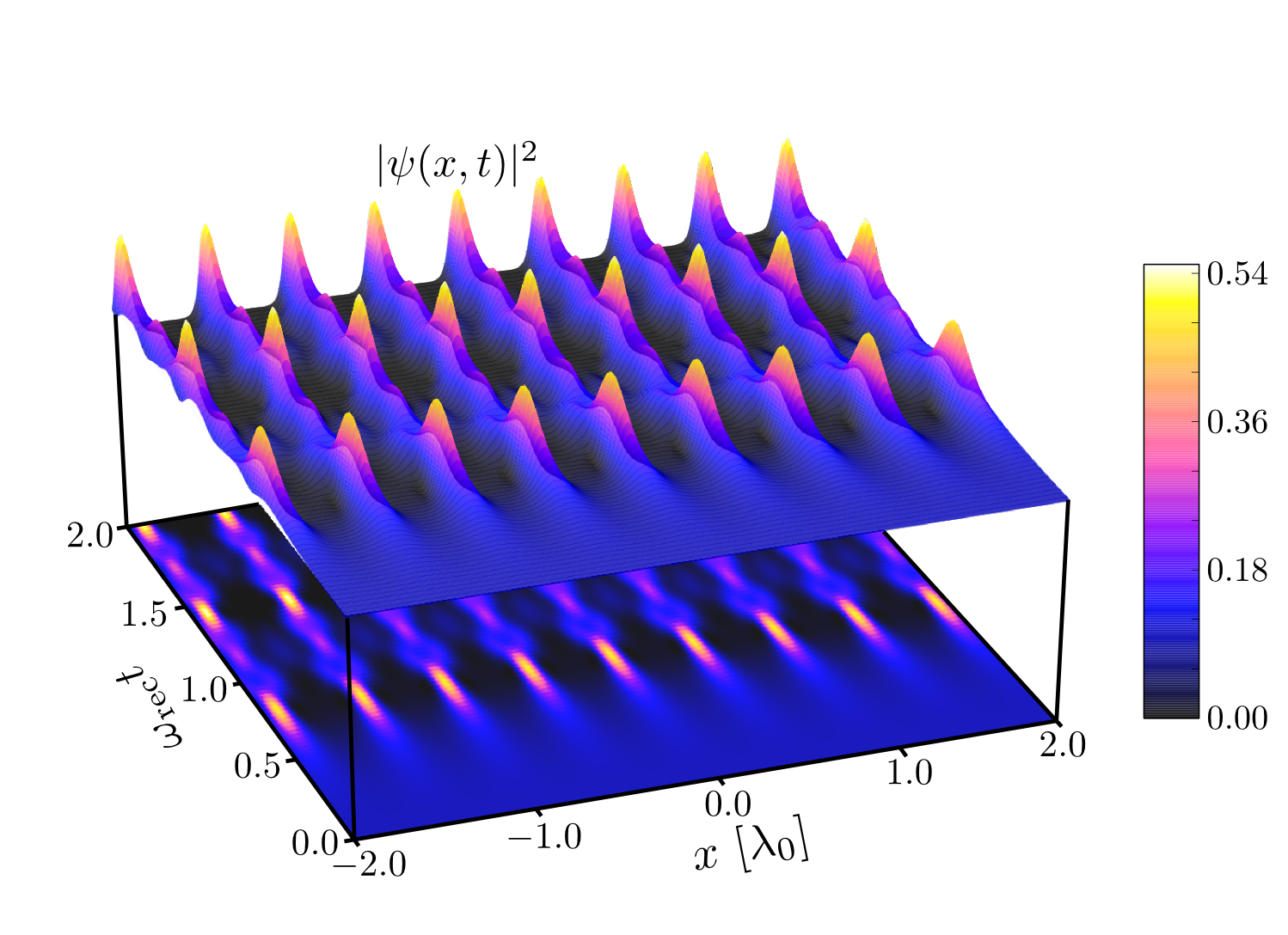}
\\
(b) \includegraphics[width=0.45\textwidth]{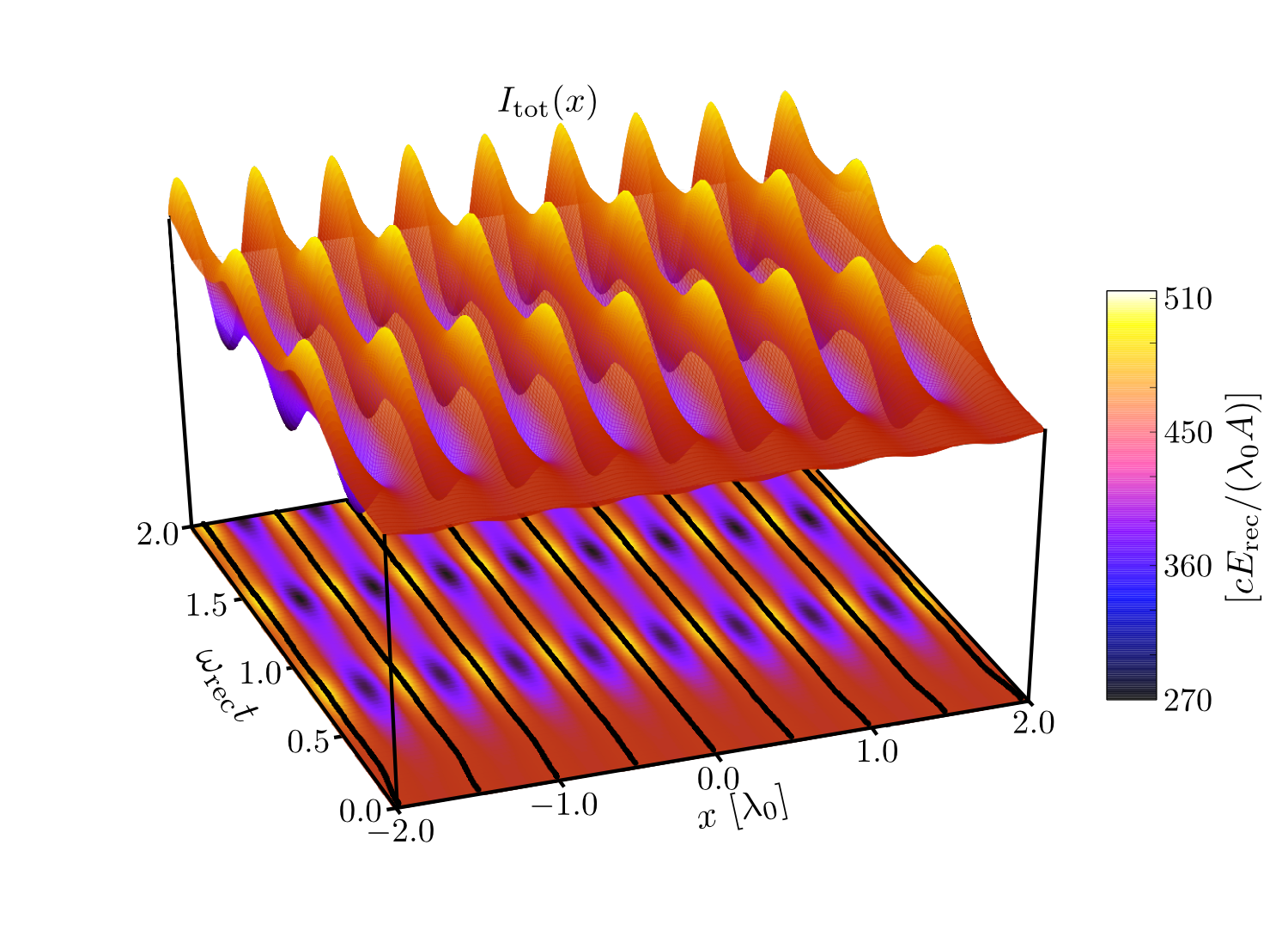}
\caption{Real time dynamics of the (a) BEC density distribution and b) the total light intensity for the same parameters as in~\fref{fig:groundstate}. The solid black lines in figure (b) show the time evolution of the intensity maxima.}
\label{fig:densint_realt}
\end{figure}

As apparent from the behavior of the reflectivity and kinetic energy
$
E_{\rm kin}(t)=\int dx\hbar^2|\partial_x\psi|^2/2m,
$
the crystalline order is reached after a few inverse recoil frequencies, after which both quantities perform oscillations about a finite value.
These residual oscillations are triggered by the energy gained by the system upon forming the density grating together with the optical lattice. The reason why this effect takes in a prominent role in the studied case is found by looking at~\fref{fig:zoom}, which shows the zoom into two peaks of the intensity distribution of the crystal.
One recognizes that the maxima of the intensity distributions of the two fields coming from left and right (blue dots in~\fref{fig:zoom}) do not coincide with the maximum of the total intensity distribution (black dot in~\fref{fig:zoom}) at which the atoms are trapped. Therefore, the trapped atoms feel a strong field gradient for each single component because they do not sit at the maxima of the two counterpropagating fields, as it would be for example the case in optical lattices. This leads to a large coupling between the two counterpropagating fields and the atoms, leading to strong long-range interactions inducing collective excitations.

\begin{figure*}
\centering
(a) \includegraphics[width=0.6\textwidth]{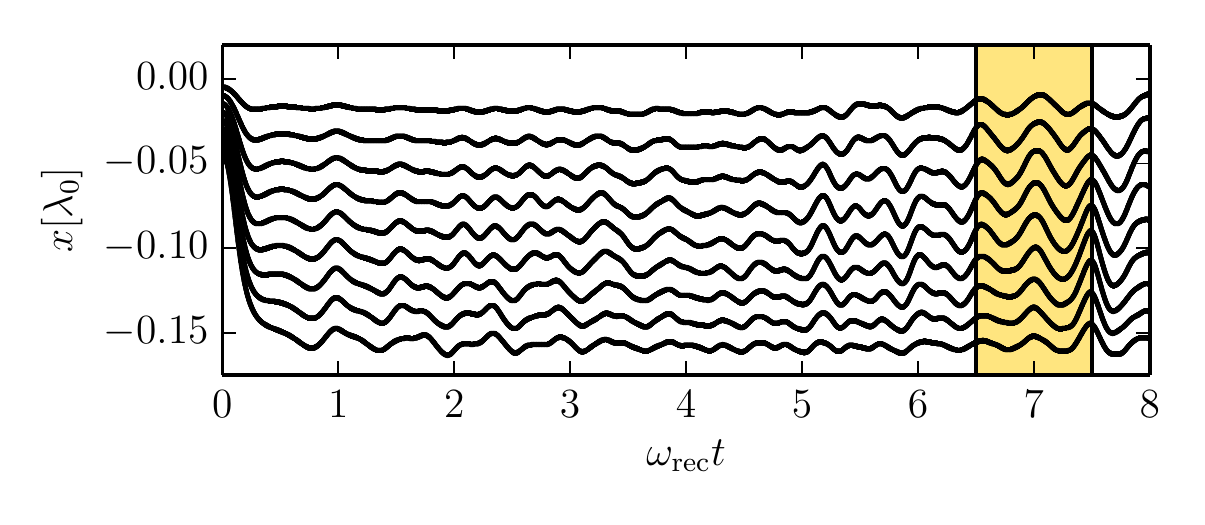}
(b) \includegraphics[width=0.3\textwidth]{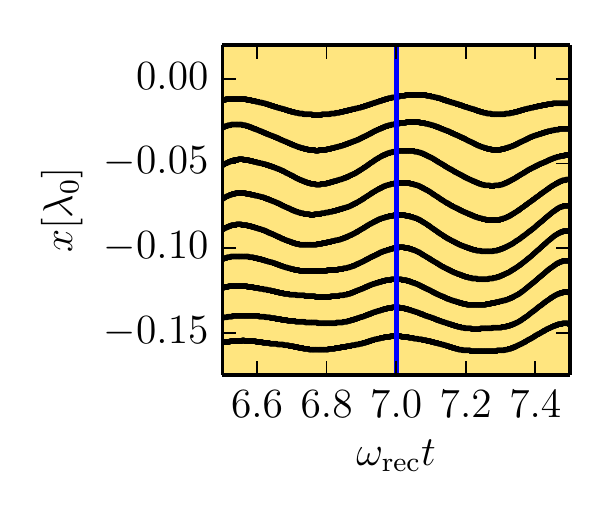}
\caption{Real time evolution of the maxima of the intensity distribution as it is shown in~\fref{fig:densint_realt}. To simplify the comparison between the single curves the maxima positions were shifted so that they all start at $x=0$. Fig. (a) shows the total time evolution where one can clearly recognize collective phonon-like excitations of the lattice after $\omega_{rec}t\gtrsim 5$. Fig. (b) shows the zoom into the yellow marked area in fig.(a) in order to demonstrate the slight dephasing between the oscillations of the maxima. All parameters are chosen as in~\fref{fig:groundstate}.}
\label{fig:phonons}
\end{figure*}

The corresponding dynamics of the BEC density and the total light intensity is shown in~\fref{fig:densint_realt}. As one can see from the solid lines marking the evolution of the intensity maxima, they start at a lattice spacing of $\lambda_0/2$ and move closer together in time reaching the emergent spacing $d$.
In addition, we see the presence of residual oscillations about the crystalline order. In particular, the light intensity shows both compression modes, modulating the amplitude of the optical lattice in time, and phonons, modulating the spacing.
The latter are clearly visible from the dynamics of the intensity maxima shown in~\fref{fig:phonons}.
Since we are neglecting retardation of the fields, the energy can be redistributed among
the collective degrees of freedom but not dissipated.
Initially, for $\omega_{rec}t\sim 1$, mostly compression modes are excited. Subsequently part of the energy stored in compression modes is transferred to lattice phonons for $\omega_{rec}t\gtrsim 5$. 
In Fig.~\ref{fig:phonons} we see a single-frequency oscillation of the intensity maxima, the latter moving almost in phase. This indeed corresponds to a low-wavelength lattice phonon, which becomes occupied for long enough times. As discussed in section~\ref{sec:crystal_excitations}, the longest wavelength is of the order of the system size $L$, consistently with the almost in-phase oscillations of Fig.~\ref{fig:phonons}.

As discussed in the previous section, lattice-phonons have a finite gap. They can efficently be excited in a quench experiment provided the energy available for collective excitations is large enough compared to $\Delta_{\rm ph}$ (see Eq.~(\ref{eq:ph_gap})).

\section{Experimental implementation with ultracold bosons} \label{sec:exp}

BECs with high densities and a controlled shape trapped in optical dipole traps are currently available in many laboratories. In principle, the setups normally employed are already very close to the one needed to study the crystallization effects presented in this work.  In the following, we will discuss the conditions needed to study our model in realistic experimental conditions, as well as the required parameter regime for observing the crystallization. Let us remark that the basic physics underlying the crystallization transition discussed here does not rely on the atoms being Bose-condensed. This phenomenon could in principle also be observed with thermal clouds or fermionic gases. Apart from the fundamentally very interesting feature of supersolidity, the practical advantage of a BEC with respect to a thermal cloud resides in its high density and low temperature, both decreasing the required laser power. On the other hand, for degenerate fermi gases, one could expect a strong dependence on the ratio between Fermi momentum and lattice constant \cite{keeling_fermi_2014,piazza_fermi_2014,zhai_fermi_2014,sandner_2015}.

We start by noting that using single beam optical traps can also lead to heating instabilities but never generate a stationary lattice~\cite{piovella2001superradiant}. Similarly, operating very close to an atomic resonance has been shown to generate instabilities and a short time formation of an optical lattice structures via so called end-fire modes~\cite{inouye1999superradiant}. As this requires significant atomic excitation, it involves fast transverse acceleration with heating and destruction of the BEC. This is prevented in our model by an improved geometry and much larger atom-field detuning.

Our model~\eqref{eqn:GPE}-\eqref{eqn:Helmh_total} is essentially 1D, which relies on the assumption that both, the atoms and the light move and propagate essentially unidirectionally along $x$. In practice this can be implemented by using a transverse trapping of the atoms tight enough to freeze out the dynamics along $y,z$. With harmonic trapping potentials this amounts to the requirement that $\omega_{y,z}^{\rm ho}$ is sufficiently larger than the BEC chemical potential $\mu$. Here we still describe the one-dimensional BEC using the GP equation, which requires the atom density to be large enough to be in the mean-field regime~\cite{string_pit}.
The enforcement of unidirectional propagation of light is more demanding since an appreciable amount of diffraction out of the BEC axis would be present inducing propagation also perpendicular to $x$. Apart from the use of hollow-core optical fibers around the BEC~\cite{christensen2008trapping}, one option available in many laboratories today is using a two-dimensional array of tubes with spacing comparable with the wavelength of the light. This arrangement would generically produce destructive interference between the transverse field components diffracted from different tubes, so that if the latter are long enough only the forward propagation along the tube axis would remain. In this configuration, each tube will act equally while the field propagates inside a medium with a refractive index given by the sum of the contributions from each tube. Indeed, since all tubes share the same backreflected field there is a natural synchronization of the different tube lattices.

\begin{figure}
\centering
(a) \includegraphics[width=0.45\textwidth]{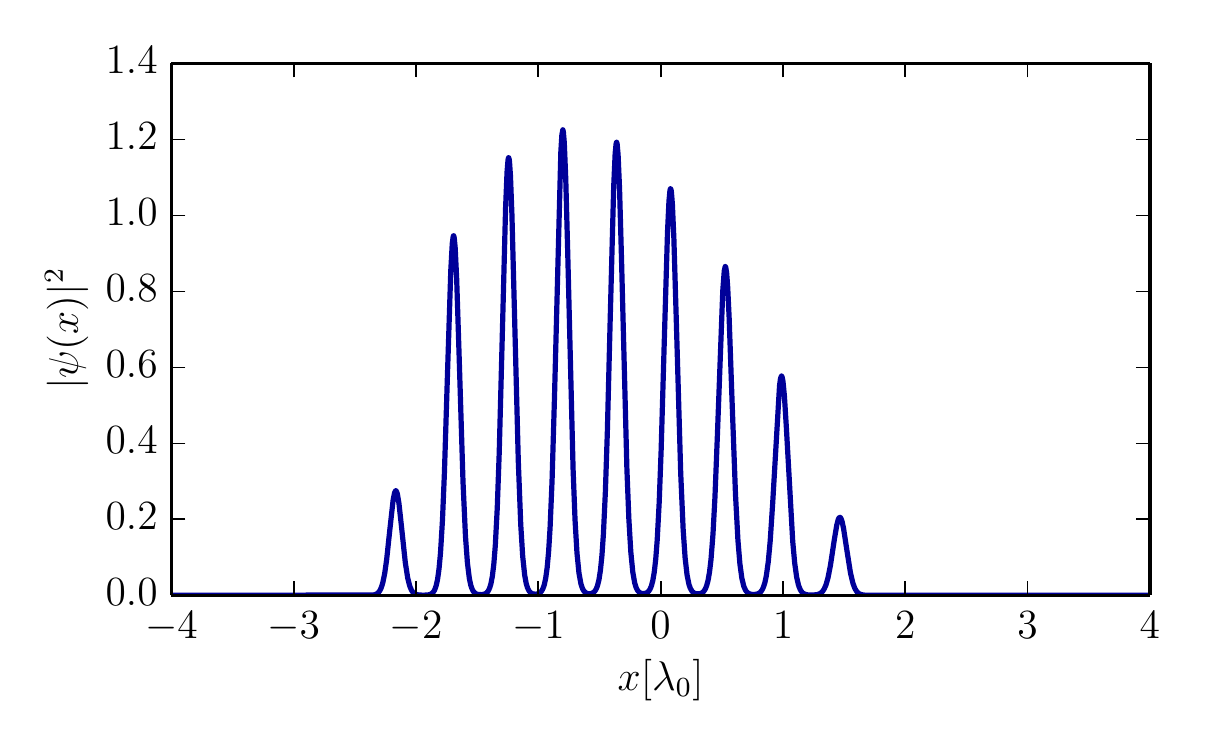}
\\
(b) \includegraphics[width=0.45\textwidth]{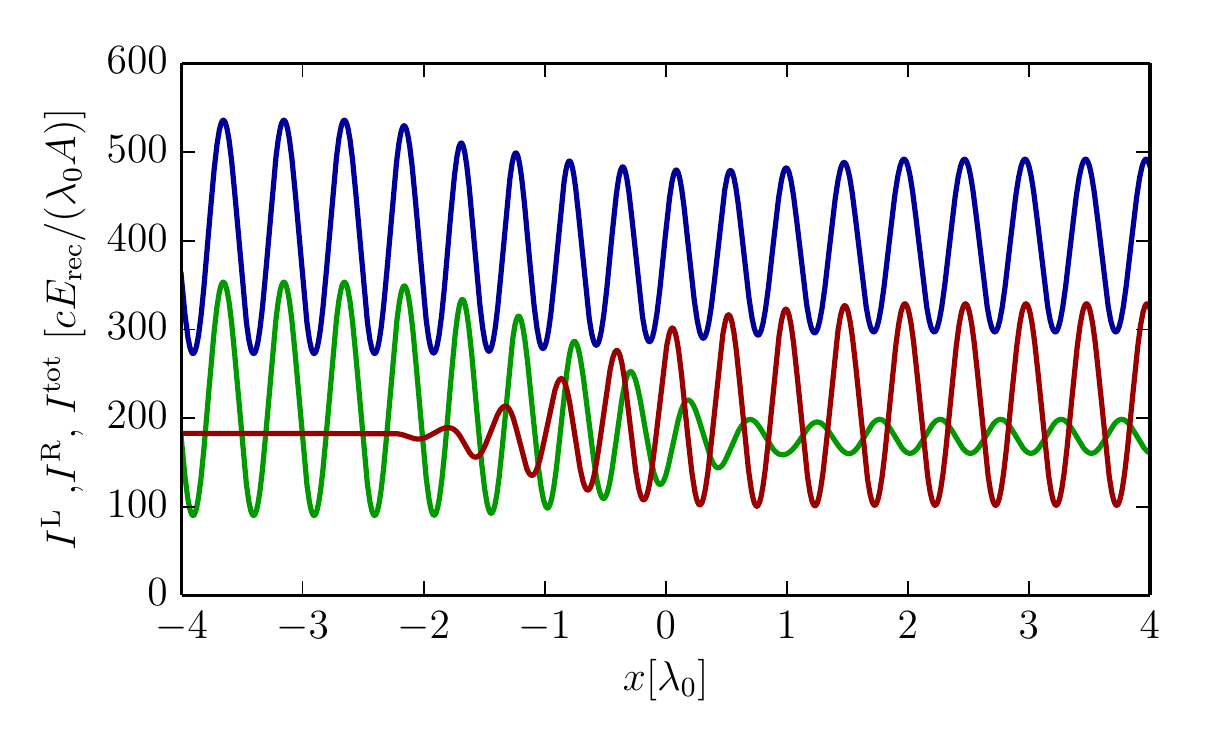}
\caption{a) Crystal ground state for and b) corresponding intensity distribution for the field from left (green) and right (red) for the same parameters as in~\fref{fig:groundstate} with an additional external potential $V_\mathrm{ext}(x)=\frac{E_\mathrm{trap}}{2}  x^2/\lambda_0^2$ with $E_\mathrm{trap}=1.0 E_\mathrm{rec}$ and for different pump intensities from left and right $I_l=200$ and $I_r=150$. The solid blue line depicts the sum of both intensities.}
\label{fig:groundstate_trap}
\end{figure}

In any experimental realization a trap to confine the BEC along $x$ will also be present. In addition, the two laser intensities might differ to some extend due to experimental inaccuracies. As an exemplary case we study the crystallization as in section~\ref{sec:groundstate} but add a harmonic trapping potential $V_\mathrm{ext}(x)=\frac{E_\mathrm{trap}}{2} x^2/\lambda_0^2$ and chose different pump intensities $I_l \neq I_r$. It can be seen from~\fref{fig:groundstate_trap} that the qualitative features of the crystalline phase remain the same as in the homogeneous case. The only difference is the parabolic envelope for the density as well as for the light intensity distribution and the shift of the distribution towards the direction of the higher intensity. The threshold behaviour remains similar to the one presented in~\fref{fig:threshold} with the only difference being an increase of the threshold intensity.
A useful feature of the considered configuration is that the crystallization process can be observed in real time by looking at the amount of reflected light, since the transmitted part of the counterpropagating beam can be separated from the reflected part having orthogonal polarization.

In order to choose the most suitable atomic transition, pump detuning and power, as well as BEC parameters like density and extension, one must consider the following constraints: we need to have i) a low enough critical driving strength (\ref{eqn:Icrit}), which depends on the detuning $\Delta_a$ and spontaneous emission $\gamma$ through the real part of polarizability $\mathrm{Re}\alpha\sim \gamma/\Delta_a$, reading 
\begin{equation}
I_{c}^{\rm L,R}\sim E_{rec}\frac{\Delta_a^2}{\gamma^2}\frac{\lambda_0}{n_A}\;,
\label{eqn:Icrit_experiment}
\end{equation}
and at the same time ii) a low enough BEC heating rate, which at the critical power reads
\begin{equation}
\Gamma_{\rm heat}\sim I_c^{\rm L,R} \frac{\gamma^2}{\Delta_a^2}\sim E_{rec}\frac{\lambda_0}{n_A}\;,
\label{eqn:heating}
\end{equation}
with $n_A=N/A$ being the surface density of the medium with respect to the light propagation.
From Eqs.~(\ref{eqn:Icrit_experiment}) and (\ref{eqn:heating}) one sees that the crystallisation is easier achieved before the BEC is heated up if we increase the BEC surface density $n_A$. There is no favorable scaling neither with detuning $\Delta_a$ nor with the linewidth $\gamma$, since both heating rate (\ref{eqn:heating}) and critical power (\ref{eqn:Icrit_experiment}) scale with $\gamma^2/\Delta_a^2$.
For commonly employed transitions like the Rb or Cs D lines, the required laser power is easily achieved, but the heating rate can become a problem at too low densities due to the required laser powers and detunings. For instance, taking $N=10^6$ atoms confined over a transverse cross section $A\sim 5\times 5\mu\text{m}^2$ and $\lambda_0\sim\mu{\rm m}$, we estimate a required power $I_c\sim \mathrm{W}/\text{cm}^2 $ with a heating rate $\Gamma_{\rm heat}\sim \text{10 Hz}$ for the Rubidium $780$nm line with a detuning $\Delta_a=100$GHz as well as for the Cesium $D2$ line with a detuning $\Delta_a=20$GHz. Such a heating rate would still allow to observe the crystal fromation since, as we see in~\fref{fig:densint_realt}, this process takes place on the inverse recoil time scale, which is of the order of ms.

\section{Conclusions and Outlook}
We predict that in suitable geometries roton instabilities originating from non-linear free space atom light interactions can be tailored to generate stationary crystalline states. They involve an optical lattice showing an emergent spacing and phononic excitations, trapping the atoms at the intensity maxima.

The required translation invariant, mirror symmetric  geometry can be realized using two orthogonal polarization degrees of freedom or frequency shifted counterpropagating beams.
We estimate that the dynamics studied in this work should be accessible in already existing experimental setups on large quasi-1D Bose-Einstein condensates. Actually, in comparison with standard crossed beam dipole traps, one simply has to adapt and control the polarizations of the trapping lasers and choose suitable detunings. The ordering process should be easily observable not only by measuring the atomic distributions but directly by looking at the reflected light from the condensate. This non-destructive measurement allows for a real-time monitoring of the dynamics.

Our results open up an intriguing new direction in quantum simulations with ultracold atoms in optical lattices, where the latter are enriched by the presence of collective phononic excitations resulting from the spontaneous crystallisation of light.
In this spirit, the application of our approach to two-dimensions and the inclusions of retardation effects as well as quantum fluctuations constitute the natural extension of this study.

\begin{acknowledgments}
We thank S. Kr\"amer for support in the numerical implementation and F. Meinert and T. Donner for helpful discussions on the experimental limitations and implementability of the system. We also thank J. Lang for useful discussions. We acknowledge support by the Austrian Science Fund FWF through projects SFB FoQuS P13 and I1697-N27. FP is supported by the APART program of the Austrian Academy of Science.
\end{acknowledgments}

\appendix

\section{Calculation of excitation spectra}\label{app:linearize}

Here we describe in detail how the linearization of the Helmholtz and the GP equation leads to the collective excitation spectra below (see~\eref{eqn:exspectr}) and above the threshold (see Fig.~\ref{fig:spect_above}).

It is convenient to slightly re-write the equations presented in section~\ref{sec:model}. Therefore, we define the relevant parameters of the system and useful units. We introduce the recoil energy $E_\mathrm{rec}:=\hbar\omega_{rec}=\hbar^2 k_0^2/(2m)$ relative to the wave number $k_0=2\pi/\lambda_0$ of the incoming lasers in vacuum. The dimensionless time is defined through the recoil frequency: $\tilde t:=\omega_\mathrm{rec}t$. The dimensionless space variable is given in units of the incoming laser wavelength $\tilde x:= x/\lambda_0$. We also rescale the fields to have units of energy: $\tilde E_\mathrm{L,R}:=\sqrt{\alpha}E_\mathrm{L,R}/\sqrt{A}$ and the atom-atom s-wave coupling to have units of energy times length: $\tilde g_\mathrm{c}:=g_\mathrm{c}/A$. 
The GP equation (\ref{eqn:GPE}) then reads
\begin{multline}
i\frac{\partial}{\partial \tilde t}\tilde\psi(\tilde x,\tilde t)=-\frac{1}{(2\pi)^2}\frac{\partial^2}{\partial \tilde x^2}\tilde{\psi}(\tilde x,\tilde t)\\
+\frac{V_\mathrm{ext}}{E_\mathrm{rec}}\tilde{\psi}(\tilde x,\tilde t)-\frac{1}{E_\mathrm{rec}}\left(|\tilde E_\mathrm{L}(\tilde x)|^2+|\tilde E_\mathrm{R}(\tilde x)|^2\right)\tilde{\psi}(\tilde x,\tilde t)\\
+\frac{\tilde g_c N}{E_\mathrm{rec}}|\tilde{\psi}(\tilde x,\tilde t)|^2\tilde{\psi}(\tilde x,\tilde t),
\label{eqn:GPEend}
\end{multline}
and the Helmholtz equations (\ref{eqn:Helmh_total}) become
\begin{equation}
\frac{\partial^2}{\partial \tilde x^2}\tilde E_\mathrm{L,R}(\tilde x)+(2\pi)^2\left[1+\zeta|\tilde \psi(\tilde x,t)|^2\right]\tilde E_\mathrm{L,R}(\tilde x)=0\;.
\label{eqn:Helmhend}
\end{equation}

Let us first consider the linearization of the Helmholtz equation~\eqref{eqn:Helmhend}. Inserting the ansatz already presented in section~\ref{sec:crys_thres} namely $\psi=(\psi_0+\delta\psi)e^{-i\mu t}$ and $E_{\mathrm{L},\mathrm{R}}=E_{L,R}^{(0)}+\delta E_{L,R}$ into~\eqref{eqn:Helmhend} and neglecting terms of second order leads to
\begin{align}
\partial_{xx}E^{(0)}_{L,R}+&(2\pi)^2\left[1+\zeta|\psi_0|^2\right]E^{(0)}_{L,R} \label{eqn:steady}\\
+\partial_{xx}\delta E_{L,R}+&(2\pi)^2\left[1+\zeta|\psi_0|^2\right]\delta E_{L,R}\label{eqn:HHperturb}\\
+&(2\pi)^2\zeta \left[\psi_0\delta \psi^*+\delta\psi\psi_0^*\right]E^{(0)}_{L,R}=0 \label{eqn:Ecoupling}.
\end{align}
The first line~\eqref{eqn:steady} corresponds to the Helmholtz equation for the steady state $E^{(0)}_\mathrm{L,R}$ an therefore it is equal to zero. The second line~\eqref{eqn:HHperturb} is the Helmholtz equation for the field perturbation whereas the third line~\eqref{eqn:Ecoupling} describes the linear coupling between the field and the BEC.

This equation can be rewritten in the following form
\begin{multline}
\left(M+K_\mathrm{eff}^2 \right)\cdot\delta\mathbf{E}_\mathrm{L,R}=\\
=-(2\pi)^2\zeta \mathcal{E}_\mathrm{L,R}^{(0)}\cdot\left(\Psi_0\cdot\delta\pmb{\psi}^*+h.c.\right)
\label{eqn:linHformal}
\end{multline}
where we defined the matrices
\begin{align}
M(x,x')&:=\partial_{xx}\delta(x-x')\\
K_\mathrm{eff}^2(x,x')&:=(2\pi)^2\left[1+\zeta n_0(x)\right]\delta(x-x')\\
\mathcal{E}_\mathrm{L,R}^{(0)}(x,x')&:=E_\mathrm{L,R}^{(0)}\delta(x-x')\\
\Psi_0(x,x')&:=\psi_0\delta(x-x')
\end{align}
and the scalar product
\begin{equation}
M\cdot\mathbf{f}= \int dx'M(x,x')f(x').
\end{equation}
The formal solution of the linearized Helmholtz equation~\eqref{eqn:linHformal} is
\begin{equation}
\delta\mathbf{E}_\mathrm{L,R}=-(2\pi)^2\zeta(M+K_\mathrm{eff}^2)^{-1}\cdot\mathcal{E}_\mathrm{L,R}^{(0)}\cdot(\Psi_0\cdot\pmb{\psi}^*+h.c.).
\label{eqn:formalsolH}
\end{equation}

The linearization of the Gross-Pitaevski equation~\eqref{eqn:GPEend} follows a similar procedure as presented above. Performing the same ansatz and neglecting the second order terms leads to
\begin{widetext}
\begin{multline}
i\partial_t \delta\psi +\mu\delta\psi=\\
-\frac{1}{(2\pi)^2}\partial_{xx}\delta\psi-\frac{1}{E_{rec}}\big[\psi_0\left(E^{(0)^*}_{L}\delta E_L+E^{(0)^*}_{R}\delta E_R+c.c.\right)+\delta\psi\left(|E^{(0)}_L|^2+|E^{(0)}_R|^2\right)\big]+\frac{g_cN}{E_{rec}}\left[|\psi_0|^2\delta\psi+\psi_0\left(\psi_0^*\delta\psi+\psi_0\delta\psi^*\right)\right]\\
-\psi_0\mu-\frac{1}{(2\pi)^2}\partial_{xx}\psi_0-\frac{1}{E_{rec}}\psi_0\left(|E^{(0)}_L|^2+|E^{(0)}_R|^2\right)+\frac{g_cN}{E_{rec}}|\psi_0|^2\psi_0
\label{eqn:GP_ansatz}.
\end{multline}
\end{widetext}
The last line of equation~\eqref{eqn:GP_ansatz} corresponds to the stationary GP equation and therefore it vanishe, as it defines how the chemical potential is related to the field amplitude and the particle particle interaction $g_c$, namely via
\begin{equation}
\mu=\frac{g_cN}{LE_{rec}}-\frac{2|E^{(0)}_\mathrm{L,R}|^2}{E_{rec}}.
\end{equation}
Inserting the formal solution~\eqref{eqn:formalsolH} into the linearized GP equation~\eqref{eqn:GP_ansatz} and performing a Fourier Transform via $f(x)=\frac{1}{\sqrt{L}}\sum_ke^{ikx}f(k)$ and $M(x,x')=\frac{1}{L}\sum_{k,k'}e^{ikx}e^{ik'x'}M(k,k')$gives
\begin{multline}
i\partial_t \mathbb{1} \pmb{\psi}=\\
\left(-\mu\mathbb{1}+T+A_L+\tilde{A}_L+A_R+\tilde{A}_R+I_\mathrm{tot}+2\nu_0\right)\pmb{\psi}+\\
+(A_L+\tilde{A}_L+A_R+\tilde{A}_R+\nu_0)\mathcal{P}\pmb{\psi}^*
\label{eqn:kGPE}
\end{multline}
where $\mathbb{1}$ denotes the identity matrix and $\mathcal{P}$ is the parity operator, \ie $\mathcal{P}\pmb{\psi}(k)=\pmb{\psi}(-k)$. We defined the following matrices:
\begin{align}
T(k,k')&:=\frac{k^2}{(2\pi)^2}\delta(k-k') \label{eqn:m1}\\
\mathcal{I}_\mathrm{tot}(k,k')&:=-\frac{1}{E_\mathrm{rec}\sqrt{L}}I_\mathrm{tot}(k-k')\\
\nu_0(k,k')&:=\frac{g_cN}{E_\mathrm{rec}\sqrt{L}}n_0(k-k')\\
A_\mathrm{L,R}(k,k')&:=\\
&\frac{\zeta\pi^2}{E_\mathrm{rec}L}\sum_{k_1,k_2}V^\dagger_\mathrm{L,R}(k,k_1)Q^{-1}(k_1,k_2)V_\mathrm{L,R}(k_2,k')\\
\tilde{A}_\mathrm{L,R}(k,k')&:=\\
&\frac{\zeta\pi^2}{E_\mathrm{rec}L}\sum_{k_1,k_2}V_\mathrm{L,R}(k,k_1)Q^{-1}(k_1,k_2)V^\dagger_\mathrm{L,R}(k_2,k')\label{eqn:m2}
\end{align}
where $I_\mathrm{tot}(k)$ and $n_0(k)$ are the Fourier transforms of the total intensity distribution and the BEC density. Besides we defined the additional matrices $V_\mathrm{L,R}(k,k'):=\sum_{k^{''}}\psi_0^*(k^{''})E_\mathrm{L,R}^{(0)}(k^{''}+k-k')$ and $Q(k,k'):=-k^2\delta(k-k')+1/\sqrt{L}k_\mathrm{eff}^2(k-k')$, where $k_\mathrm{eff}^2(k)$ is the Fourier transform of $(2\pi)^2\left[1+\zeta n_0(x)\right]$. In the following we will call the sum of the $A$-matrices $\mathcal{A}(k,k'):=A_L(k,k')+\tilde{A}_L(k,k')+A_R(k,k')+\tilde{A}_R(k,k')$

Let us now define the spinor $\Psi(q):=(\psi(q),\psi^*(q))^T$ where $\psi(q)$ defines a single momentum component of $\pmb{\psi}$ from~\eqref{eqn:kGPE}. This definition allows us to write the GP equation in the form $i\partial_t\Psi(q)=\sum_{q'}R(q,q')\Psi(q')$ where the matrix $R$ is defined as follows
\begin{widetext}
\begin{equation}
R(q,q')=\begin{pmatrix} 
  -\mu\delta(q-q')+\frac{q^2}{(2\pi)^2}\delta(q-q')\mathcal{I}_\mathrm{tot}(q-q')+\mathcal{A}(q,q')     & \nu_0(q,q')+\mathcal{A}(q,q')\\ 
 -\nu_0(q,q')-\mathcal{A}(-q,-q') & -\left[-\mu\delta(q-q')+\frac{q^2}{(2\pi)^2}\delta(q-q')\mathcal{I}_\mathrm{tot}(q-q')+\mathcal{A}(q,q')\right]
\end{pmatrix}.
\label{eqn:Rmat}
\end{equation}
\end{widetext}
This equation now enables us to calculate the excitation spectrum of the considered system for any arbitrary intensity and BEC density distribution by calculating the eigenvalues of the matrix $R$.
\subsection{Collective spectrum in the homogeneous phase}
If we now use the ansatz already presented in section~\ref{sec:crys_thres} namely $\psi_0(x,t)=1/\sqrt{L}$  and $E_{\rm L,R}^{(0)}=C \exp(\pm i k_{\rm eff}x)$ we can calculate the excitation spectrum below threshold. This ansatz implies $I_\mathrm{tot}(x)=|E_R^{(0)}|^2+|E_L^{(0)}|^2=2|C|^2$ and $n_0(x)=1/L$ which results in
\begin{align}
\mathcal{I}_\mathrm{tot}(k)=-\frac{8|C|^2}{E_\mathrm{rec}}\delta(k)\\
\nu_0(k)=\frac{g_cN}{E_\mathrm{rec}L}\delta(k).
\end{align}
In addition the matrices $Q$ and $V$ amount to
\begin{align}
Q(k)&=(k_\mathrm{eff}^2-k^2)\delta(k)\\
V_\mathrm{L}(k)&=C\sqrt{L}\delta(k+k_\mathrm{eff}),
\end{align}
resulting in
\begin{equation}
A_\mathrm{L,R}(q,q')=-\frac{\zeta}{E_\mathrm{rec}}\frac{|C|^2(2\pi)^2}{L}\frac{1}{q^2\mp2k_\mathrm{eff}q}\delta(q-q').
\end{equation}
Note that in this special case $\tilde{A}_\mathrm{L,R}=A_\mathrm{L,R}$.
If one now calculates the matrix $R$ via~\eqref{eqn:Rmat} and solves $\det\left[R(q-q')-\omega\mathbb{1}\right]=0$ one gets
\begin{equation}
\omega^2-\frac{q^2}{(2\pi)^2}\left[\frac{q^2}{2m}+2\frac{g_cN}{E_\mathrm{rec}L}-\frac{8\zeta}{E_\mathrm{rec}}\frac{|C|^2(2\pi)^2}{L}\frac{1}{q^2-4 k_\mathrm{eff}^2}\right]=0.
\end{equation}
Transforming this equation back into the original units leads to the excitation spectrum~\eqref{eqn:exspectr} presented in section~\ref{sec:crys_thres}.

\subsection{Collective spectrum above threshold}
Let us now move on to the calculation of the collective excitation spectrum above threshold as it is presented in section~\ref{sec:crystal_excitations}. In this case an analytical answer like the one presented in the previous section is not possible, since the translation-invariance is broken so that the matrices describing the linear system are not diagonal in momentum space. Therefore, a numerical approach is required, involving in general the discretization of the position(momentum) continuum.

The matrices defined in Eqns.~\eqref{eqn:m1}-\eqref{eqn:m2} can be calculated by numerically finding the fourier transforms of the stationary states found via complex time evolution in section~\ref{sec:groundstate}. The resulting total matrix $R$ can then be diagonalized numerically.

A further difficulty arising in our setup is that in the stationary crystalline solution, the total light intensity and atom density are periodi, whereas the intensity of each polarization component is not. This originates from the repeated reflection from the density grating, introducing the decaying evelope shown in Fig.(5b) of the main text. This prevents the use of the quasi-momentum to label the excitation modes. Therefore we use the momentum corresponding to the largest component of the eigenvector in order to order the eigenvalues in Fig.~\ref{fig:spect_above}.

\section{Numerical methods}\label{app:numerics}
The model described in section~\ref{sec:model} constitutes a scoupled system of equations~\eqref{eqn:GPE} and~\eqref{eqn:Helmh_total}. In this appendix we will shortly discuss the numerical methods we used to simulate the time evolution of the studied system as it is used in sections~\ref{sec:groundstate}-~\ref{sec:exp}.

The algorithm consists of two parts. First we need to solve the Helmholtz equation~\eqref{eqn:Helmh_total} for a given space dependent susceptibility~\eqref{eqn:susc}. This corresponds to a initial value problem with the boundary conditions
\begin{align}
E(x=-L/2)&=A_L+B_L, \label{eqn:E}\\
\partial_x E(x=-L/2)&=ik_0(A_L-BL).\label{eqn:dE}
\end{align}
Here $A_L$ and $B_L$ define the incoming ($A_L$) and outgoing ($B_L$) field amplitudes at the left side of the BEC. They are related to the amplitudes on the right side via
\begin{align}
B_L&=R A_L+T D_R \label{eqn:BL}\\
C_R&=T A_l+R D_R \label{eqn:CR}
\end{align}
with the system's reflection and transmission coefficients $R$ and $T$. Of course, these reflection and transmission coefficients depend on the system's susceptibility. They can easily be estimated by solving the HH equation for an arbitrary initial condition~\eqref{eqn:E} and~\eqref{eqn:dE}, leading to well defined fields at the boundaries allowing for an estimation of the right hand amplitudes $C_R$ and $D_R$. Hence, $R$ and $T$ can be calculated via~\eqref{eqn:BL} and~\eqref{eqn:CR}. As soon as we know the initial conditions we can find the solution of the Helmholtz equation via spatial integration performed by a fourth order Runge-Kutta solver.

The solution of the HH equation is then used to calculate the optical potential~\eqref{eqn:Vopt}. The time evolution of the GP equation with the newly found potential is then calculated by using a split step method. Note that the HH equation has to be solved within each time step resulting in a modified potential for the next time step in the GP equation. The time evolution is finished as soon as the system is found in a stationary state.

%


\begin{thebibliography}{51}%
\makeatletter
\providecommand \@ifxundefined [1]{%
 \@ifx{#1\undefined}
}%
\providecommand \@ifnum [1]{%
 \ifnum #1\expandafter \@firstoftwo
 \else \expandafter \@secondoftwo
 \fi
}%
\providecommand \@ifx [1]{%
 \ifx #1\expandafter \@firstoftwo
 \else \expandafter \@secondoftwo
 \fi
}%
\providecommand \natexlab [1]{#1}%
\providecommand \enquote  [1]{``#1''}%
\providecommand \bibnamefont  [1]{#1}%
\providecommand \bibfnamefont [1]{#1}%
\providecommand \citenamefont [1]{#1}%
\providecommand \href@noop [0]{\@secondoftwo}%
\providecommand \href [0]{\begingroup \@sanitize@url \@href}%
\providecommand \@href[1]{\@@startlink{#1}\@@href}%
\providecommand \@@href[1]{\endgroup#1\@@endlink}%
\providecommand \@sanitize@url [0]{\catcode `\\12\catcode `\$12\catcode
  `\&12\catcode `\#12\catcode `\^12\catcode `\_12\catcode `\%12\relax}%
\providecommand \@@startlink[1]{}%
\providecommand \@@endlink[0]{}%
\providecommand \url  [0]{\begingroup\@sanitize@url \@url }%
\providecommand \@url [1]{\endgroup\@href {#1}{\urlprefix }}%
\providecommand \urlprefix  [0]{URL }%
\providecommand \Eprint [0]{\href }%
\providecommand \doibase [0]{http://dx.doi.org/}%
\providecommand \selectlanguage [0]{\@gobble}%
\providecommand \bibinfo  [0]{\@secondoftwo}%
\providecommand \bibfield  [0]{\@secondoftwo}%
\providecommand \translation [1]{[#1]}%
\providecommand \BibitemOpen [0]{}%
\providecommand \bibitemStop [0]{}%
\providecommand \bibitemNoStop [0]{.\EOS\space}%
\providecommand \EOS [0]{\spacefactor3000\relax}%
\providecommand \BibitemShut  [1]{\csname bibitem#1\endcsname}%
\let\auto@bib@innerbib\@empty
\bibitem [{\citenamefont {Andreev}\ \emph {et~al.}(1980)\citenamefont
  {Andreev}, \citenamefont {Emel'yanov},\ and\ \citenamefont
  {Il'inski}}]{andreev1980collective}%
  \BibitemOpen
  \bibfield  {author} {\bibinfo {author} {\bibfnamefont {Anatolii~Vasil'evich}\
  \bibnamefont {Andreev}}, \bibinfo {author} {\bibfnamefont
  {Vladimir~Il'ich{\u\i}}\ \bibnamefont {Emel'yanov}}, \ and\ \bibinfo {author}
  {\bibfnamefont {Yu~A}\ \bibnamefont {Il'inski}},\ }\bibfield  {title}
  {\enquote {\bibinfo {title} {Collective spontaneous emission (dicke
  superradiance)},}\ }\href@noop {} {\bibfield  {journal} {\bibinfo  {journal}
  {Phys. Usp.}\ }\textbf {\bibinfo {volume} {23}},\ \bibinfo {pages} {493--514}
  (\bibinfo {year} {1980})}\BibitemShut {NoStop}%
\bibitem [{\citenamefont {Mekhov}\ \emph {et~al.}(2007)\citenamefont {Mekhov},
  \citenamefont {Maschler},\ and\ \citenamefont {Ritsch}}]{mekhov2007light}%
  \BibitemOpen
  \bibfield  {author} {\bibinfo {author} {\bibfnamefont {Igor~B}\ \bibnamefont
  {Mekhov}}, \bibinfo {author} {\bibfnamefont {Christoph}\ \bibnamefont
  {Maschler}}, \ and\ \bibinfo {author} {\bibfnamefont {Helmut}\ \bibnamefont
  {Ritsch}},\ }\bibfield  {title} {\enquote {\bibinfo {title} {Light scattering
  from ultracold atoms in optical lattices as an optical probe of quantum
  statistics},}\ }\href@noop {} {\bibfield  {journal} {\bibinfo  {journal}
  {Phys. Rev. A}\ }\textbf {\bibinfo {volume} {76}},\ \bibinfo {pages} {053618}
  (\bibinfo {year} {2007})}\BibitemShut {NoStop}%
\bibitem [{\citenamefont {Zoubi}\ and\ \citenamefont
  {Ritsch}(2010)}]{zoubi2010metastability}%
  \BibitemOpen
  \bibfield  {author} {\bibinfo {author} {\bibfnamefont {H}~\bibnamefont
  {Zoubi}}\ and\ \bibinfo {author} {\bibfnamefont {H}~\bibnamefont {Ritsch}},\
  }\bibfield  {title} {\enquote {\bibinfo {title} {Metastability and
  directional emission characteristics of excitons in 1d optical lattices},}\
  }\href@noop {} {\bibfield  {journal} {\bibinfo  {journal} {Europhys. Lett.}\
  }\textbf {\bibinfo {volume} {90}},\ \bibinfo {pages} {23001} (\bibinfo {year}
  {2010})}\BibitemShut {NoStop}%
\bibitem [{\citenamefont {Braun}\ \emph {et~al.}(1995)\citenamefont {Braun},
  \citenamefont {Korn}, \citenamefont {Liu}, \citenamefont {Du}, \citenamefont
  {Squier},\ and\ \citenamefont {Mourou}}]{braun1995self}%
  \BibitemOpen
  \bibfield  {author} {\bibinfo {author} {\bibfnamefont {A}~\bibnamefont
  {Braun}}, \bibinfo {author} {\bibfnamefont {G}~\bibnamefont {Korn}}, \bibinfo
  {author} {\bibfnamefont {X}~\bibnamefont {Liu}}, \bibinfo {author}
  {\bibfnamefont {D}~\bibnamefont {Du}}, \bibinfo {author} {\bibfnamefont
  {J}~\bibnamefont {Squier}}, \ and\ \bibinfo {author} {\bibfnamefont
  {G}~\bibnamefont {Mourou}},\ }\bibfield  {title} {\enquote {\bibinfo {title}
  {Self-channeling of high-peak-power femtosecond laser pulses in air},}\
  }\href@noop {} {\bibfield  {journal} {\bibinfo  {journal} {Opt. Lett.}\
  }\textbf {\bibinfo {volume} {20}},\ \bibinfo {pages} {73--75} (\bibinfo
  {year} {1995})}\BibitemShut {NoStop}%
\bibitem [{\citenamefont {Kartashov}\ \emph {et~al.}(2013)\citenamefont
  {Kartashov}, \citenamefont {Ali{\v{s}}auskas}, \citenamefont {Pug{\v{z}}lys},
  \citenamefont {Voronin}, \citenamefont {Zheltikov}, \citenamefont {Petrarca},
  \citenamefont {B{\'e}jot}, \citenamefont {Kasparian}, \citenamefont {Wolf},\
  and\ \citenamefont {Baltu{\v{s}}ka}}]{kartashov2013mid}%
  \BibitemOpen
  \bibfield  {author} {\bibinfo {author} {\bibfnamefont {D}~\bibnamefont
  {Kartashov}}, \bibinfo {author} {\bibfnamefont {S}~\bibnamefont
  {Ali{\v{s}}auskas}}, \bibinfo {author} {\bibfnamefont {A}~\bibnamefont
  {Pug{\v{z}}lys}}, \bibinfo {author} {\bibfnamefont {A}~\bibnamefont
  {Voronin}}, \bibinfo {author} {\bibfnamefont {A}~\bibnamefont {Zheltikov}},
  \bibinfo {author} {\bibfnamefont {Massimo}\ \bibnamefont {Petrarca}},
  \bibinfo {author} {\bibfnamefont {P}~\bibnamefont {B{\'e}jot}}, \bibinfo
  {author} {\bibfnamefont {J{\'e}r{\^o}me}\ \bibnamefont {Kasparian}}, \bibinfo
  {author} {\bibfnamefont {J-P}\ \bibnamefont {Wolf}}, \ and\ \bibinfo {author}
  {\bibfnamefont {A}~\bibnamefont {Baltu{\v{s}}ka}},\ }\bibfield  {title}
  {\enquote {\bibinfo {title} {Mid-infrared laser filamentation in molecular
  gases},}\ }\href@noop {} {\bibfield  {journal} {\bibinfo  {journal} {Opt.
  Lett.}\ }\textbf {\bibinfo {volume} {38}},\ \bibinfo {pages} {3194--3197}
  (\bibinfo {year} {2013})}\BibitemShut {NoStop}%
\bibitem [{\citenamefont {Zheltikov}\ \emph {et~al.}(2012)\citenamefont
  {Zheltikov}, \citenamefont {LHuillier},\ and\ \citenamefont
  {Krausz}}]{zheltikov2012nonlinear}%
  \BibitemOpen
  \bibfield  {author} {\bibinfo {author} {\bibfnamefont {Aleksei}\ \bibnamefont
  {Zheltikov}}, \bibinfo {author} {\bibfnamefont {Anne}\ \bibnamefont
  {LHuillier}}, \ and\ \bibinfo {author} {\bibfnamefont {Ferenc}\ \bibnamefont
  {Krausz}},\ }\bibfield  {title} {\enquote {\bibinfo {title} {Nonlinear
  optics},}\ }in\ \href@noop {} {\emph {\bibinfo {booktitle} {Springer Handbook
  of Lasers and Optics}}}\ (\bibinfo  {publisher} {Springer},\ \bibinfo {year}
  {2012})\ pp.\ \bibinfo {pages} {161--251}\BibitemShut {NoStop}%
\bibitem [{\citenamefont {Singer}\ \emph {et~al.}(2003)\citenamefont {Singer},
  \citenamefont {Frick}, \citenamefont {Bernet},\ and\ \citenamefont
  {Ritsch-Marte}}]{singer2003self}%
  \BibitemOpen
  \bibfield  {author} {\bibinfo {author} {\bibfnamefont {Wolfgang}\
  \bibnamefont {Singer}}, \bibinfo {author} {\bibfnamefont {Manfred}\
  \bibnamefont {Frick}}, \bibinfo {author} {\bibfnamefont {Stefan}\
  \bibnamefont {Bernet}}, \ and\ \bibinfo {author} {\bibfnamefont {Monika}\
  \bibnamefont {Ritsch-Marte}},\ }\bibfield  {title} {\enquote {\bibinfo
  {title} {Self-organized array of regularly spaced microbeads in a
  fiber-optical trap},}\ }\href@noop {} {\bibfield  {journal} {\bibinfo
  {journal} {J. Opt. Soc. Am. B}\ }\textbf {\bibinfo {volume} {20}},\ \bibinfo
  {pages} {1568--1574} (\bibinfo {year} {2003})}\BibitemShut {NoStop}%
\bibitem [{\citenamefont {Burns}\ \emph {et~al.}(1990)\citenamefont {Burns},
  \citenamefont {Fournier},\ and\ \citenamefont
  {Golovchenko}}]{burns1990optical}%
  \BibitemOpen
  \bibfield  {author} {\bibinfo {author} {\bibfnamefont {Michael~M}\
  \bibnamefont {Burns}}, \bibinfo {author} {\bibfnamefont {Jean-Marc}\
  \bibnamefont {Fournier}}, \ and\ \bibinfo {author} {\bibfnamefont {Jene~A}\
  \bibnamefont {Golovchenko}},\ }\bibfield  {title} {\enquote {\bibinfo {title}
  {Optical matter: crystallization and binding in intense optical fields},}\
  }\href@noop {} {\bibfield  {journal} {\bibinfo  {journal} {Science}\ }\textbf
  {\bibinfo {volume} {249}},\ \bibinfo {pages} {749} (\bibinfo {year}
  {1990})}\BibitemShut {NoStop}%
\bibitem [{\citenamefont {Tatarkova}\ \emph {et~al.}(2002)\citenamefont
  {Tatarkova}, \citenamefont {Carruthers},\ and\ \citenamefont
  {Dholakia}}]{tatarkova2002one}%
  \BibitemOpen
  \bibfield  {author} {\bibinfo {author} {\bibfnamefont {SA}~\bibnamefont
  {Tatarkova}}, \bibinfo {author} {\bibfnamefont {AE}~\bibnamefont
  {Carruthers}}, \ and\ \bibinfo {author} {\bibfnamefont {K}~\bibnamefont
  {Dholakia}},\ }\bibfield  {title} {\enquote {\bibinfo {title}
  {One-dimensional optically bound arrays of microscopic particles},}\
  }\href@noop {} {\bibfield  {journal} {\bibinfo  {journal} {Phys. Rev. Lett.}\
  }\textbf {\bibinfo {volume} {89}},\ \bibinfo {pages} {283901} (\bibinfo
  {year} {2002})}\BibitemShut {NoStop}%
\bibitem [{\citenamefont {Demergis}\ and\ \citenamefont
  {Florin}(2012)}]{demergis2012ultrastrong}%
  \BibitemOpen
  \bibfield  {author} {\bibinfo {author} {\bibfnamefont {Vassili}\ \bibnamefont
  {Demergis}}\ and\ \bibinfo {author} {\bibfnamefont {Ernst-Ludwig}\
  \bibnamefont {Florin}},\ }\bibfield  {title} {\enquote {\bibinfo {title}
  {Ultrastrong optical binding of metallic nanoparticles},}\ }\href@noop {}
  {\bibfield  {journal} {\bibinfo  {journal} {Nano Lett.}\ }\textbf {\bibinfo
  {volume} {12}},\ \bibinfo {pages} {5756--5760} (\bibinfo {year}
  {2012})}\BibitemShut {NoStop}%
\bibitem [{\citenamefont {Kar{\'a}sek}\ \emph {et~al.}(2008)\citenamefont
  {Kar{\'a}sek}, \citenamefont {{\v{C}}i{\v{z}}m{\'a}r}, \citenamefont
  {Brzobohat{\`y}}, \citenamefont {Zem{\'a}nek}, \citenamefont
  {Garc{\'e}s-Ch{\'a}vez},\ and\ \citenamefont {Dholakia}}]{karasek2008long}%
  \BibitemOpen
  \bibfield  {author} {\bibinfo {author} {\bibfnamefont {V}~\bibnamefont
  {Kar{\'a}sek}}, \bibinfo {author} {\bibfnamefont {T}~\bibnamefont
  {{\v{C}}i{\v{z}}m{\'a}r}}, \bibinfo {author} {\bibfnamefont {O}~\bibnamefont
  {Brzobohat{\`y}}}, \bibinfo {author} {\bibfnamefont {P}~\bibnamefont
  {Zem{\'a}nek}}, \bibinfo {author} {\bibfnamefont {V}~\bibnamefont
  {Garc{\'e}s-Ch{\'a}vez}}, \ and\ \bibinfo {author} {\bibfnamefont
  {K}~\bibnamefont {Dholakia}},\ }\bibfield  {title} {\enquote {\bibinfo
  {title} {Long-range one-dimensional longitudinal optical binding},}\
  }\href@noop {} {\bibfield  {journal} {\bibinfo  {journal} {Phys. Rev. Lett.}\
  }\textbf {\bibinfo {volume} {101}},\ \bibinfo {pages} {143601} (\bibinfo
  {year} {2008})}\BibitemShut {NoStop}%
\bibitem [{\citenamefont {Kar{\'a}sek}\ \emph {et~al.}(2009)\citenamefont
  {Kar{\'a}sek}, \citenamefont {Brzobohat{\`y}},\ and\ \citenamefont
  {Zem{\'a}nek}}]{karasek2009longitudinal}%
  \BibitemOpen
  \bibfield  {author} {\bibinfo {author} {\bibfnamefont {V}~\bibnamefont
  {Kar{\'a}sek}}, \bibinfo {author} {\bibfnamefont {O}~\bibnamefont
  {Brzobohat{\`y}}}, \ and\ \bibinfo {author} {\bibfnamefont {P}~\bibnamefont
  {Zem{\'a}nek}},\ }\bibfield  {title} {\enquote {\bibinfo {title}
  {Longitudinal optical binding of several spherical particles studied by the
  coupled dipole method},}\ }\href@noop {} {\bibfield  {journal} {\bibinfo
  {journal} {J. Opt. A}\ }\textbf {\bibinfo {volume} {11}},\ \bibinfo {pages}
  {034009} (\bibinfo {year} {2009})}\BibitemShut {NoStop}%
\bibitem [{\citenamefont {Dholakia}\ and\ \citenamefont
  {Zem{\'a}nek}(2010)}]{dholakia2010colloquium}%
  \BibitemOpen
  \bibfield  {author} {\bibinfo {author} {\bibfnamefont {Kishan}\ \bibnamefont
  {Dholakia}}\ and\ \bibinfo {author} {\bibfnamefont {Pavel}\ \bibnamefont
  {Zem{\'a}nek}},\ }\bibfield  {title} {\enquote {\bibinfo {title} {Colloquium:
  Gripped by light: Optical binding},}\ }\href@noop {} {\bibfield  {journal}
  {\bibinfo  {journal} {Rev. Mod. Phys.}\ }\textbf {\bibinfo {volume} {82}},\
  \bibinfo {pages} {1767} (\bibinfo {year} {2010})}\BibitemShut {NoStop}%
\bibitem [{\citenamefont {Bonifacio}\ and\ \citenamefont
  {De~Salvo}(1994)}]{bonifacio1994collective}%
  \BibitemOpen
  \bibfield  {author} {\bibinfo {author} {\bibfnamefont {R}~\bibnamefont
  {Bonifacio}}\ and\ \bibinfo {author} {\bibfnamefont {L}~\bibnamefont
  {De~Salvo}},\ }\bibfield  {title} {\enquote {\bibinfo {title} {Collective
  atomic recoil laser (carl) optical gain without inversion by collective
  atomic recoil and self-bunching of two-level atoms},}\ }\href@noop {}
  {\bibfield  {journal} {\bibinfo  {journal} {Nuclear Instruments and Methods
  in Physics Research Section A: Accelerators, Spectrometers, Detectors and
  Associated Equipment}\ }\textbf {\bibinfo {volume} {341}},\ \bibinfo {pages}
  {360--362} (\bibinfo {year} {1994})}\BibitemShut {NoStop}%
\bibitem [{\citenamefont {Saffman}(1998)}]{saffman1998self}%
  \BibitemOpen
  \bibfield  {author} {\bibinfo {author} {\bibfnamefont {M}~\bibnamefont
  {Saffman}},\ }\bibfield  {title} {\enquote {\bibinfo {title} {Self-induced
  dipole force and filamentation instability of a matter wave},}\ }\href@noop
  {} {\bibfield  {journal} {\bibinfo  {journal} {Phys. Rev. Lett.}\ }\textbf
  {\bibinfo {volume} {81}},\ \bibinfo {pages} {65} (\bibinfo {year}
  {1998})}\BibitemShut {NoStop}%
\bibitem [{\citenamefont {Inouye}\ \emph {et~al.}(1999)\citenamefont {Inouye},
  \citenamefont {Chikkatur}, \citenamefont {Stamper-Kurn}, \citenamefont
  {Stenger}, \citenamefont {Pritchard},\ and\ \citenamefont
  {Ketterle}}]{inouye1999superradiant}%
  \BibitemOpen
  \bibfield  {author} {\bibinfo {author} {\bibfnamefont {S}~\bibnamefont
  {Inouye}}, \bibinfo {author} {\bibfnamefont {AP}~\bibnamefont {Chikkatur}},
  \bibinfo {author} {\bibfnamefont {DM}~\bibnamefont {Stamper-Kurn}}, \bibinfo
  {author} {\bibfnamefont {J}~\bibnamefont {Stenger}}, \bibinfo {author}
  {\bibfnamefont {DE}~\bibnamefont {Pritchard}}, \ and\ \bibinfo {author}
  {\bibfnamefont {W}~\bibnamefont {Ketterle}},\ }\bibfield  {title} {\enquote
  {\bibinfo {title} {Superradiant rayleigh scattering from a bose-einstein
  condensate},}\ }\href@noop {} {\bibfield  {journal} {\bibinfo  {journal}
  {Science}\ }\textbf {\bibinfo {volume} {285}},\ \bibinfo {pages} {571--574}
  (\bibinfo {year} {1999})}\BibitemShut {NoStop}%
\bibitem [{\citenamefont {Piovella}\ \emph {et~al.}(2001)\citenamefont
  {Piovella}, \citenamefont {Bonifacio}, \citenamefont {McNeil},\ and\
  \citenamefont {Robb}}]{piovella2001superradiant}%
  \BibitemOpen
  \bibfield  {author} {\bibinfo {author} {\bibfnamefont {N}~\bibnamefont
  {Piovella}}, \bibinfo {author} {\bibfnamefont {R}~\bibnamefont {Bonifacio}},
  \bibinfo {author} {\bibfnamefont {BWJ}\ \bibnamefont {McNeil}}, \ and\
  \bibinfo {author} {\bibfnamefont {GRM}\ \bibnamefont {Robb}},\ }\bibfield
  {title} {\enquote {\bibinfo {title} {Superradiant light scattering and
  grating formation in cold atomic vapours},}\ }\href@noop {} {\bibfield
  {journal} {\bibinfo  {journal} {Opt. Commun.}\ }\textbf {\bibinfo {volume}
  {187}},\ \bibinfo {pages} {165--170} (\bibinfo {year} {2001})}\BibitemShut
  {NoStop}%
\bibitem [{\citenamefont {O’dell}\ \emph {et~al.}(2003)\citenamefont
  {O’dell}, \citenamefont {Giovanazzi},\ and\ \citenamefont
  {Kurizki}}]{o2003rotons}%
  \BibitemOpen
  \bibfield  {author} {\bibinfo {author} {\bibfnamefont {DHJ}\ \bibnamefont
  {O’dell}}, \bibinfo {author} {\bibfnamefont {S}~\bibnamefont {Giovanazzi}},
  \ and\ \bibinfo {author} {\bibfnamefont {G}~\bibnamefont {Kurizki}},\
  }\bibfield  {title} {\enquote {\bibinfo {title} {Rotons in gaseous
  bose-einstein condensates irradiated by a laser},}\ }\href@noop {} {\bibfield
   {journal} {\bibinfo  {journal} {Phys. Rev. Lett.}\ }\textbf {\bibinfo
  {volume} {90}},\ \bibinfo {pages} {110402} (\bibinfo {year}
  {2003})}\BibitemShut {NoStop}%
\bibitem [{\citenamefont {Yoshikawa}\ \emph {et~al.}(2005)\citenamefont
  {Yoshikawa}, \citenamefont {Torii},\ and\ \citenamefont {Kuga}}]{kuga_2005}%
  \BibitemOpen
  \bibfield  {author} {\bibinfo {author} {\bibfnamefont {Yutaka}\ \bibnamefont
  {Yoshikawa}}, \bibinfo {author} {\bibfnamefont {Yoshio}\ \bibnamefont
  {Torii}}, \ and\ \bibinfo {author} {\bibfnamefont {Takahiro}\ \bibnamefont
  {Kuga}},\ }\bibfield  {title} {\enquote {\bibinfo {title} {Superradiant light
  scattering from thermal atomic vapors},}\ }\href@noop {} {\bibfield
  {journal} {\bibinfo  {journal} {Phys. Rev. Lett.}\ }\textbf {\bibinfo
  {volume} {94}},\ \bibinfo {pages} {083602} (\bibinfo {year}
  {2005})}\BibitemShut {NoStop}%
\bibitem [{\citenamefont {Muradyan}\ \emph {et~al.}(2005)\citenamefont
  {Muradyan}, \citenamefont {Wang}, \citenamefont {Williams},\ and\
  \citenamefont {Saffman}}]{muradyan2005absolute}%
  \BibitemOpen
  \bibfield  {author} {\bibinfo {author} {\bibfnamefont {Gevorg~A}\
  \bibnamefont {Muradyan}}, \bibinfo {author} {\bibfnamefont {Yingxue}\
  \bibnamefont {Wang}}, \bibinfo {author} {\bibfnamefont {William}\
  \bibnamefont {Williams}}, \ and\ \bibinfo {author} {\bibfnamefont {Mark}\
  \bibnamefont {Saffman}},\ }\bibfield  {title} {\enquote {\bibinfo {title}
  {Absolute instability and pattern formation in cold atomic vapors},}\ }in\
  \href@noop {} {\emph {\bibinfo {booktitle} {Nonlinear Guided Waves and Their
  Applications}}}\ (\bibinfo {organization} {Optical Society of America},\
  \bibinfo {year} {2005})\BibitemShut {NoStop}%
\bibitem [{\citenamefont {Slama}\ \emph {et~al.}(2007)\citenamefont {Slama},
  \citenamefont {Bux}, \citenamefont {Krenz}, \citenamefont {Zimmermann},\ and\
  \citenamefont {Courteille}}]{zimmermann_2007}%
  \BibitemOpen
  \bibfield  {author} {\bibinfo {author} {\bibfnamefont {S.}~\bibnamefont
  {Slama}}, \bibinfo {author} {\bibfnamefont {S.}~\bibnamefont {Bux}}, \bibinfo
  {author} {\bibfnamefont {G.}~\bibnamefont {Krenz}}, \bibinfo {author}
  {\bibfnamefont {C.}~\bibnamefont {Zimmermann}}, \ and\ \bibinfo {author}
  {\bibfnamefont {Ph.~W.}\ \bibnamefont {Courteille}},\ }\bibfield  {title}
  {\enquote {\bibinfo {title} {Superradiant rayleigh scattering and collective
  atomic recoil lasing in a ring cavity},}\ }\href@noop {} {\bibfield
  {journal} {\bibinfo  {journal} {Phys. Rev. Lett.}\ }\textbf {\bibinfo
  {volume} {98}},\ \bibinfo {pages} {053603} (\bibinfo {year}
  {2007})}\BibitemShut {NoStop}%
\bibitem [{\citenamefont {Greenberg}\ \emph {et~al.}(2011)\citenamefont
  {Greenberg}, \citenamefont {Schmittberger},\ and\ \citenamefont
  {Gauthier}}]{greenberg2011bunching}%
  \BibitemOpen
  \bibfield  {author} {\bibinfo {author} {\bibfnamefont {Joel~A}\ \bibnamefont
  {Greenberg}}, \bibinfo {author} {\bibfnamefont {Bonnie~L}\ \bibnamefont
  {Schmittberger}}, \ and\ \bibinfo {author} {\bibfnamefont {Daniel~J}\
  \bibnamefont {Gauthier}},\ }\bibfield  {title} {\enquote {\bibinfo {title}
  {Bunching-induced optical nonlinearity and instability in cold atoms
  [invited]},}\ }\href@noop {} {\bibfield  {journal} {\bibinfo  {journal} {Opt.
  Express}\ }\textbf {\bibinfo {volume} {19}},\ \bibinfo {pages} {22535--22549}
  (\bibinfo {year} {2011})}\BibitemShut {NoStop}%
\bibitem [{\citenamefont {Schmittberger}\ and\ \citenamefont
  {Gauthier}(2016)}]{schmittberger2016spontaneous}%
  \BibitemOpen
  \bibfield  {author} {\bibinfo {author} {\bibfnamefont {Bonnie~L}\
  \bibnamefont {Schmittberger}}\ and\ \bibinfo {author} {\bibfnamefont
  {Daniel~J}\ \bibnamefont {Gauthier}},\ }\bibfield  {title} {\enquote
  {\bibinfo {title} {Spontaneous emergence of free-space optical and atomic
  patterns},}\ }\href@noop {} {\bibfield  {journal} {\bibinfo  {journal} {arXiv
  preprint arXiv:1603.06294}\ } (\bibinfo {year} {2016})}\BibitemShut {NoStop}%
\bibitem [{\citenamefont {Labeyrie}\ \emph {et~al.}(2014)\citenamefont
  {Labeyrie}, \citenamefont {Tesio}, \citenamefont {Gomes}, \citenamefont
  {Oppo}, \citenamefont {Firth}, \citenamefont {Robb}, \citenamefont {Arnold},
  \citenamefont {Kaiser},\ and\ \citenamefont
  {Ackemann}}]{labeyrie2014optomechanical}%
  \BibitemOpen
  \bibfield  {author} {\bibinfo {author} {\bibfnamefont {Guillaume}\
  \bibnamefont {Labeyrie}}, \bibinfo {author} {\bibfnamefont {Enrico}\
  \bibnamefont {Tesio}}, \bibinfo {author} {\bibfnamefont {Pedro~M}\
  \bibnamefont {Gomes}}, \bibinfo {author} {\bibfnamefont {G-L}\ \bibnamefont
  {Oppo}}, \bibinfo {author} {\bibfnamefont {William~J}\ \bibnamefont {Firth}},
  \bibinfo {author} {\bibfnamefont {Gordon~RM}\ \bibnamefont {Robb}}, \bibinfo
  {author} {\bibfnamefont {Aidan~S}\ \bibnamefont {Arnold}}, \bibinfo {author}
  {\bibfnamefont {Robin}\ \bibnamefont {Kaiser}}, \ and\ \bibinfo {author}
  {\bibfnamefont {Thorsten}\ \bibnamefont {Ackemann}},\ }\bibfield  {title}
  {\enquote {\bibinfo {title} {Optomechanical self-structuring in a cold atomic
  gas},}\ }\href@noop {} {\bibfield  {journal} {\bibinfo  {journal} {Nature
  Photon.}\ }\textbf {\bibinfo {volume} {8}},\ \bibinfo {pages} {321--325}
  (\bibinfo {year} {2014})}\BibitemShut {NoStop}%
\bibitem [{\citenamefont {Robb}\ \emph {et~al.}(2015)\citenamefont {Robb},
  \citenamefont {Tesio}, \citenamefont {Oppo}, \citenamefont {Firth},
  \citenamefont {Ackemann},\ and\ \citenamefont {Bonifacio}}]{robb_BEC_2015}%
  \BibitemOpen
  \bibfield  {author} {\bibinfo {author} {\bibfnamefont {G.~R.~M.}\
  \bibnamefont {Robb}}, \bibinfo {author} {\bibfnamefont {E.}~\bibnamefont
  {Tesio}}, \bibinfo {author} {\bibfnamefont {G.-L.}\ \bibnamefont {Oppo}},
  \bibinfo {author} {\bibfnamefont {W.~J.}\ \bibnamefont {Firth}}, \bibinfo
  {author} {\bibfnamefont {T.}~\bibnamefont {Ackemann}}, \ and\ \bibinfo
  {author} {\bibfnamefont {R.}~\bibnamefont {Bonifacio}},\ }\bibfield  {title}
  {\enquote {\bibinfo {title} {Quantum threshold for optomechanical
  self-structuring in a bose-einstein condensate},}\ }\href@noop {} {\bibfield
  {journal} {\bibinfo  {journal} {Phys. Rev. Lett.}\ }\textbf {\bibinfo
  {volume} {114}},\ \bibinfo {pages} {173903} (\bibinfo {year}
  {2015})}\BibitemShut {NoStop}%
\bibitem [{\citenamefont {Domokos}\ and\ \citenamefont
  {Ritsch}(2002)}]{domokos2002collective}%
  \BibitemOpen
  \bibfield  {author} {\bibinfo {author} {\bibfnamefont {Peter}\ \bibnamefont
  {Domokos}}\ and\ \bibinfo {author} {\bibfnamefont {Helmut}\ \bibnamefont
  {Ritsch}},\ }\bibfield  {title} {\enquote {\bibinfo {title} {Collective
  cooling and self-organization of atoms in a cavity},}\ }\href@noop {}
  {\bibfield  {journal} {\bibinfo  {journal} {Phys. Rev. Lett.}\ }\textbf
  {\bibinfo {volume} {89}},\ \bibinfo {pages} {253003} (\bibinfo {year}
  {2002})}\BibitemShut {NoStop}%
\bibitem [{\citenamefont {Black}\ \emph {et~al.}(2003)\citenamefont {Black},
  \citenamefont {Chan},\ and\ \citenamefont {Vuleti\ifmmode~\acute{c}\else
  \'{c}\fi{}}}]{vuletic_2003}%
  \BibitemOpen
  \bibfield  {author} {\bibinfo {author} {\bibfnamefont {Adam~T.}\ \bibnamefont
  {Black}}, \bibinfo {author} {\bibfnamefont {Hilton~W.}\ \bibnamefont {Chan}},
  \ and\ \bibinfo {author} {\bibfnamefont {Vladan}\ \bibnamefont
  {Vuleti\ifmmode~\acute{c}\else \'{c}\fi{}}},\ }\bibfield  {title} {\enquote
  {\bibinfo {title} {Observation of collective friction forces due to spatial
  self-organization of atoms: From rayleigh to bragg scattering},}\ }\href@noop
  {} {\bibfield  {journal} {\bibinfo  {journal} {Phys. Rev. Lett.}\ }\textbf
  {\bibinfo {volume} {91}},\ \bibinfo {pages} {203001} (\bibinfo {year}
  {2003})}\BibitemShut {NoStop}%
\bibitem [{\citenamefont {Arnold}\ \emph {et~al.}(2012)\citenamefont {Arnold},
  \citenamefont {Baden},\ and\ \citenamefont {Barrett}}]{barrett_2012}%
  \BibitemOpen
  \bibfield  {author} {\bibinfo {author} {\bibfnamefont {K.~J.}\ \bibnamefont
  {Arnold}}, \bibinfo {author} {\bibfnamefont {M.~P.}\ \bibnamefont {Baden}}, \
  and\ \bibinfo {author} {\bibfnamefont {M.~D.}\ \bibnamefont {Barrett}},\
  }\bibfield  {title} {\enquote {\bibinfo {title} {Self-organization threshold
  scaling for thermal atoms coupled to a cavity},}\ }\href@noop {} {\bibfield
  {journal} {\bibinfo  {journal} {Phys. Rev. Lett.}\ }\textbf {\bibinfo
  {volume} {109}},\ \bibinfo {pages} {153002} (\bibinfo {year}
  {2012})}\BibitemShut {NoStop}%
\bibitem [{\citenamefont {Ritsch}\ \emph {et~al.}(2013)\citenamefont {Ritsch},
  \citenamefont {Domokos}, \citenamefont {Brennecke},\ and\ \citenamefont
  {Esslinger}}]{ritsch2013cold}%
  \BibitemOpen
  \bibfield  {author} {\bibinfo {author} {\bibfnamefont {Helmut}\ \bibnamefont
  {Ritsch}}, \bibinfo {author} {\bibfnamefont {Peter}\ \bibnamefont {Domokos}},
  \bibinfo {author} {\bibfnamefont {Ferdinand}\ \bibnamefont {Brennecke}}, \
  and\ \bibinfo {author} {\bibfnamefont {Tilman}\ \bibnamefont {Esslinger}},\
  }\bibfield  {title} {\enquote {\bibinfo {title} {Cold atoms in
  cavity-generated dynamical optical potentials},}\ }\href@noop {} {\bibfield
  {journal} {\bibinfo  {journal} {Rev. Mod. Phys.}\ }\textbf {\bibinfo {volume}
  {85}},\ \bibinfo {pages} {553} (\bibinfo {year} {2013})}\BibitemShut
  {NoStop}%
\bibitem [{\citenamefont {Baumann}\ \emph {et~al.}(2010)\citenamefont
  {Baumann}, \citenamefont {Guerlin}, \citenamefont {Brennecke},\ and\
  \citenamefont {Esslinger}}]{baumann2010dicke}%
  \BibitemOpen
  \bibfield  {author} {\bibinfo {author} {\bibfnamefont {Kristian}\
  \bibnamefont {Baumann}}, \bibinfo {author} {\bibfnamefont {Christine}\
  \bibnamefont {Guerlin}}, \bibinfo {author} {\bibfnamefont {Ferdinand}\
  \bibnamefont {Brennecke}}, \ and\ \bibinfo {author} {\bibfnamefont {Tilman}\
  \bibnamefont {Esslinger}},\ }\bibfield  {title} {\enquote {\bibinfo {title}
  {Dicke quantum phase transition with a superfluid gas in an optical
  cavity},}\ }\href@noop {} {\bibfield  {journal} {\bibinfo  {journal}
  {Nature}\ }\textbf {\bibinfo {volume} {464}},\ \bibinfo {pages} {1301--1306}
  (\bibinfo {year} {2010})}\BibitemShut {NoStop}%
\bibitem [{\citenamefont {Ke{\ss}ler}\ \emph {et~al.}(2014)\citenamefont
  {Ke{\ss}ler}, \citenamefont {Klinder}, \citenamefont {Wolke},\ and\
  \citenamefont {Hemmerich}}]{kessler2014steering}%
  \BibitemOpen
  \bibfield  {author} {\bibinfo {author} {\bibfnamefont {H}~\bibnamefont
  {Ke{\ss}ler}}, \bibinfo {author} {\bibfnamefont {J}~\bibnamefont {Klinder}},
  \bibinfo {author} {\bibfnamefont {M}~\bibnamefont {Wolke}}, \ and\ \bibinfo
  {author} {\bibfnamefont {A}~\bibnamefont {Hemmerich}},\ }\bibfield  {title}
  {\enquote {\bibinfo {title} {Steering matter wave superradiance with an
  ultranarrow-band optical cavity},}\ }\href@noop {} {\bibfield  {journal}
  {\bibinfo  {journal} {Phys. Rev. Lett.}\ }\textbf {\bibinfo {volume} {113}},\
  \bibinfo {pages} {070404} (\bibinfo {year} {2014})}\BibitemShut {NoStop}%
\bibitem [{\citenamefont {Ostermann}\ \emph {et~al.}(2015)\citenamefont
  {Ostermann}, \citenamefont {Grie{\ss}er},\ and\ \citenamefont
  {Ritsch}}]{ostermann2015atomic}%
  \BibitemOpen
  \bibfield  {author} {\bibinfo {author} {\bibfnamefont {S}~\bibnamefont
  {Ostermann}}, \bibinfo {author} {\bibfnamefont {T}~\bibnamefont
  {Grie{\ss}er}}, \ and\ \bibinfo {author} {\bibfnamefont {H}~\bibnamefont
  {Ritsch}},\ }\bibfield  {title} {\enquote {\bibinfo {title} {Atomic
  self-ordering in a ring cavity with counterpropagating pump fields},}\
  }\href@noop {} {\bibfield  {journal} {\bibinfo  {journal} {Europhys. Lett.)}\
  }\textbf {\bibinfo {volume} {109}},\ \bibinfo {pages} {43001} (\bibinfo
  {year} {2015})}\BibitemShut {NoStop}%
\bibitem [{\citenamefont {Mottl}\ \emph
  {et~al.}(2012{\natexlab{a}})\citenamefont {Mottl}, \citenamefont {Brennecke},
  \citenamefont {Baumann}, \citenamefont {Landig}, \citenamefont {Donner},\
  and\ \citenamefont {Esslinger}}]{mottl2012roton}%
  \BibitemOpen
  \bibfield  {author} {\bibinfo {author} {\bibfnamefont {R}~\bibnamefont
  {Mottl}}, \bibinfo {author} {\bibfnamefont {F}~\bibnamefont {Brennecke}},
  \bibinfo {author} {\bibfnamefont {K}~\bibnamefont {Baumann}}, \bibinfo
  {author} {\bibfnamefont {R}~\bibnamefont {Landig}}, \bibinfo {author}
  {\bibfnamefont {T}~\bibnamefont {Donner}}, \ and\ \bibinfo {author}
  {\bibfnamefont {T}~\bibnamefont {Esslinger}},\ }\bibfield  {title} {\enquote
  {\bibinfo {title} {Roton-type mode softening in a quantum gas with
  cavity-mediated long-range interactions},}\ }\href@noop {} {\bibfield
  {journal} {\bibinfo  {journal} {Science}\ }\textbf {\bibinfo {volume}
  {336}},\ \bibinfo {pages} {1570--1573} (\bibinfo {year}
  {2012}{\natexlab{a}})}\BibitemShut {NoStop}%
\bibitem [{\citenamefont {Grie{\ss}er}\ and\ \citenamefont
  {Ritsch}(2013)}]{griesser2013light}%
  \BibitemOpen
  \bibfield  {author} {\bibinfo {author} {\bibfnamefont {Tobias}\ \bibnamefont
  {Grie{\ss}er}}\ and\ \bibinfo {author} {\bibfnamefont {Helmut}\ \bibnamefont
  {Ritsch}},\ }\bibfield  {title} {\enquote {\bibinfo {title} {Light-induced
  crystallization of cold atoms in a 1d optical trap},}\ }\href@noop {}
  {\bibfield  {journal} {\bibinfo  {journal} {Phys. Rev. Lett.}\ }\textbf
  {\bibinfo {volume} {111}},\ \bibinfo {pages} {055702} (\bibinfo {year}
  {2013})}\BibitemShut {NoStop}%
\bibitem [{\citenamefont {Chang}\ \emph {et~al.}(2013)\citenamefont {Chang},
  \citenamefont {Cirac},\ and\ \citenamefont {Kimble}}]{chang2013self}%
  \BibitemOpen
  \bibfield  {author} {\bibinfo {author} {\bibfnamefont {DE}~\bibnamefont
  {Chang}}, \bibinfo {author} {\bibfnamefont {J~Ignacio}\ \bibnamefont
  {Cirac}}, \ and\ \bibinfo {author} {\bibfnamefont {HJ}~\bibnamefont
  {Kimble}},\ }\bibfield  {title} {\enquote {\bibinfo {title}
  {Self-organization of atoms along a nanophotonic waveguide},}\ }\href@noop {}
  {\bibfield  {journal} {\bibinfo  {journal} {Phys. Rev. Lett.}\ }\textbf
  {\bibinfo {volume} {110}},\ \bibinfo {pages} {113606} (\bibinfo {year}
  {2013})}\BibitemShut {NoStop}%
\bibitem [{\citenamefont {Gopalakrishnan}\ \emph {et~al.}(2009)\citenamefont
  {Gopalakrishnan}, \citenamefont {Lev},\ and\ \citenamefont
  {Goldbart}}]{gopalakrishnan2009emergent}%
  \BibitemOpen
  \bibfield  {author} {\bibinfo {author} {\bibfnamefont {Sarang}\ \bibnamefont
  {Gopalakrishnan}}, \bibinfo {author} {\bibfnamefont {Benjamin~L}\
  \bibnamefont {Lev}}, \ and\ \bibinfo {author} {\bibfnamefont {Paul~M}\
  \bibnamefont {Goldbart}},\ }\bibfield  {title} {\enquote {\bibinfo {title}
  {Emergent crystallinity and frustration with bose--einstein condensates in
  multimode cavities},}\ }\href@noop {} {\bibfield  {journal} {\bibinfo
  {journal} {Nat. Phys.}\ }\textbf {\bibinfo {volume} {5}},\ \bibinfo {pages}
  {845--850} (\bibinfo {year} {2009})}\BibitemShut {NoStop}%
\bibitem [{\citenamefont {Chang}\ \emph {et~al.}(2008)\citenamefont {Chang},
  \citenamefont {Gritsev}, \citenamefont {Morigi}, \citenamefont {Vuleti{\'c}},
  \citenamefont {Lukin},\ and\ \citenamefont
  {Demler}}]{chang2008crystallization}%
  \BibitemOpen
  \bibfield  {author} {\bibinfo {author} {\bibfnamefont {DE}~\bibnamefont
  {Chang}}, \bibinfo {author} {\bibfnamefont {V}~\bibnamefont {Gritsev}},
  \bibinfo {author} {\bibfnamefont {G}~\bibnamefont {Morigi}}, \bibinfo
  {author} {\bibfnamefont {V}~\bibnamefont {Vuleti{\'c}}}, \bibinfo {author}
  {\bibfnamefont {MD}~\bibnamefont {Lukin}}, \ and\ \bibinfo {author}
  {\bibfnamefont {EA}~\bibnamefont {Demler}},\ }\bibfield  {title} {\enquote
  {\bibinfo {title} {Crystallization of strongly interacting photons in a
  nonlinear optical fibre},}\ }\href@noop {} {\bibfield  {journal} {\bibinfo
  {journal} {Nat. Phys.}\ }\textbf {\bibinfo {volume} {4}},\ \bibinfo {pages}
  {884--889} (\bibinfo {year} {2008})}\BibitemShut {NoStop}%
\bibitem [{\citenamefont {Otterbach}\ \emph {et~al.}(2013)\citenamefont
  {Otterbach}, \citenamefont {Moos}, \citenamefont {Muth},\ and\ \citenamefont
  {Fleischhauer}}]{otterbach_2013}%
  \BibitemOpen
  \bibfield  {author} {\bibinfo {author} {\bibfnamefont {Johannes}\
  \bibnamefont {Otterbach}}, \bibinfo {author} {\bibfnamefont {Matthias}\
  \bibnamefont {Moos}}, \bibinfo {author} {\bibfnamefont {Dominik}\
  \bibnamefont {Muth}}, \ and\ \bibinfo {author} {\bibfnamefont {Michael}\
  \bibnamefont {Fleischhauer}},\ }\bibfield  {title} {\enquote {\bibinfo
  {title} {Wigner crystallization of single photons in cold rydberg
  ensembles},}\ }\href@noop {} {\bibfield  {journal} {\bibinfo  {journal}
  {Phys. Rev. Lett.}\ }\textbf {\bibinfo {volume} {111}},\ \bibinfo {pages}
  {113001} (\bibinfo {year} {2013})}\BibitemShut {NoStop}%
\bibitem [{\citenamefont {Stringari}\ and\ \citenamefont
  {Pitaevskii}(2003)}]{string_pit}%
  \BibitemOpen
  \bibfield  {author} {\bibinfo {author} {\bibfnamefont {S.}~\bibnamefont
  {Stringari}}\ and\ \bibinfo {author} {\bibfnamefont {L.}~\bibnamefont
  {Pitaevskii}},\ }\href@noop {} {\emph {\bibinfo {title} {Bose-Einstein
  Condensation}}}\ (\bibinfo  {publisher} {Oxford University Press},\ \bibinfo
  {year} {2003})\BibitemShut {NoStop}%
\bibitem [{\citenamefont {Chaikin}\ and\ \citenamefont
  {Lubensky}(2000)}]{chaikin_lub}%
  \BibitemOpen
  \bibfield  {author} {\bibinfo {author} {\bibfnamefont {P.M.}\ \bibnamefont
  {Chaikin}}\ and\ \bibinfo {author} {\bibfnamefont {T.C.}\ \bibnamefont
  {Lubensky}},\ }\href@noop {} {\emph {\bibinfo {title} {Principles of
  Condensed Matter Physics}}}\ (\bibinfo  {publisher} {Cambridge University
  Press},\ \bibinfo {year} {2000})\BibitemShut {NoStop}%
\bibitem [{\citenamefont {Mottl}\ \emph
  {et~al.}(2012{\natexlab{b}})\citenamefont {Mottl}, \citenamefont {Brennecke},
  \citenamefont {Baumann}, \citenamefont {Landig}, \citenamefont {Donner},\
  and\ \citenamefont {Esslinger}}]{eth_soft}%
  \BibitemOpen
  \bibfield  {author} {\bibinfo {author} {\bibfnamefont {R.}~\bibnamefont
  {Mottl}}, \bibinfo {author} {\bibfnamefont {F.}~\bibnamefont {Brennecke}},
  \bibinfo {author} {\bibfnamefont {K.}~\bibnamefont {Baumann}}, \bibinfo
  {author} {\bibfnamefont {R.}~\bibnamefont {Landig}}, \bibinfo {author}
  {\bibfnamefont {T.}~\bibnamefont {Donner}}, \ and\ \bibinfo {author}
  {\bibfnamefont {T.}~\bibnamefont {Esslinger}},\ }\bibfield  {title} {\enquote
  {\bibinfo {title} {Rotontype mode softening in a quantum gas with
  cavity-mediated long-range interactions},}\ }\href@noop {} {\bibfield
  {journal} {\bibinfo  {journal} {Science}\ }\textbf {\bibinfo {volume}
  {336}},\ \bibinfo {pages} {1570--1573} (\bibinfo {year}
  {2012}{\natexlab{b}})}\BibitemShut {NoStop}%
\bibitem [{\citenamefont {Ostermann}\ \emph {et~al.}(2014)\citenamefont
  {Ostermann}, \citenamefont {Sonnleitner},\ and\ \citenamefont
  {Ritsch}}]{ostermann2014scattering}%
  \BibitemOpen
  \bibfield  {author} {\bibinfo {author} {\bibfnamefont {Stefan}\ \bibnamefont
  {Ostermann}}, \bibinfo {author} {\bibfnamefont {Matthias}\ \bibnamefont
  {Sonnleitner}}, \ and\ \bibinfo {author} {\bibfnamefont {Helmut}\
  \bibnamefont {Ritsch}},\ }\bibfield  {title} {\enquote {\bibinfo {title}
  {Scattering approach to two-colour light forces and self-ordering of
  polarizable particles},}\ }\href@noop {} {\bibfield  {journal} {\bibinfo
  {journal} {New J. Phys.}\ }\textbf {\bibinfo {volume} {16}},\ \bibinfo
  {pages} {043017} (\bibinfo {year} {2014})}\BibitemShut {NoStop}%
\bibitem [{\citenamefont {Deutsch}\ \emph {et~al.}(1995)\citenamefont
  {Deutsch}, \citenamefont {Spreeuw}, \citenamefont {Rolston},\ and\
  \citenamefont {Phillips}}]{deutsch1995photonic}%
  \BibitemOpen
  \bibfield  {author} {\bibinfo {author} {\bibfnamefont {IH}~\bibnamefont
  {Deutsch}}, \bibinfo {author} {\bibfnamefont {RJC}\ \bibnamefont {Spreeuw}},
  \bibinfo {author} {\bibfnamefont {SL}~\bibnamefont {Rolston}}, \ and\
  \bibinfo {author} {\bibfnamefont {WD}~\bibnamefont {Phillips}},\ }\bibfield
  {title} {\enquote {\bibinfo {title} {Photonic band gaps in optical
  lattices},}\ }\href@noop {} {\bibfield  {journal} {\bibinfo  {journal} {Phys.
  Rev. A}\ }\textbf {\bibinfo {volume} {52}},\ \bibinfo {pages} {1394}
  (\bibinfo {year} {1995})}\BibitemShut {NoStop}%
\bibitem [{\citenamefont {Ashcroft}\ and\ \citenamefont
  {Mermin}(1976)}]{ashcroft1976solid}%
  \BibitemOpen
  \bibfield  {author} {\bibinfo {author} {\bibfnamefont {Neil~W}\ \bibnamefont
  {Ashcroft}}\ and\ \bibinfo {author} {\bibfnamefont {N~David}\ \bibnamefont
  {Mermin}},\ }\bibfield  {title} {\enquote {\bibinfo {title} {Solid state
  phys},}\ }\href@noop {} {\bibfield  {journal} {\bibinfo  {journal} {Saunders,
  Philadelphia}\ }\textbf {\bibinfo {volume} {293}} (\bibinfo {year}
  {1976})}\BibitemShut {NoStop}%
\bibitem [{\citenamefont {Wigner}(1938)}]{wigner1938effects}%
  \BibitemOpen
  \bibfield  {author} {\bibinfo {author} {\bibfnamefont {Eugene}\ \bibnamefont
  {Wigner}},\ }\bibfield  {title} {\enquote {\bibinfo {title} {Effects of the
  electron interaction on the energy levels of electrons in metals},}\
  }\href@noop {} {\bibfield  {journal} {\bibinfo  {journal} {Transactions of
  the Faraday Society}\ }\textbf {\bibinfo {volume} {34}},\ \bibinfo {pages}
  {678--685} (\bibinfo {year} {1938})}\BibitemShut {NoStop}%
\bibitem [{\citenamefont {Yukalov}\ and\ \citenamefont
  {Ziegler}(2015)}]{yukalov_2015}%
  \BibitemOpen
  \bibfield  {author} {\bibinfo {author} {\bibfnamefont {V.~I.}\ \bibnamefont
  {Yukalov}}\ and\ \bibinfo {author} {\bibfnamefont {K.}~\bibnamefont
  {Ziegler}},\ }\bibfield  {title} {\enquote {\bibinfo {title} {Instability of
  insulating states in optical lattices due to collective phonon
  excitations},}\ }\href {\doibase 10.1103/PhysRevA.91.023628} {\bibfield
  {journal} {\bibinfo  {journal} {Phys. Rev. A}\ }\textbf {\bibinfo {volume}
  {91}},\ \bibinfo {pages} {023628} (\bibinfo {year} {2015})}\BibitemShut
  {NoStop}%
\bibitem [{\citenamefont {Keeling}\ \emph {et~al.}(2014)\citenamefont
  {Keeling}, \citenamefont {Bhaseen},\ and\ \citenamefont
  {Simons}}]{keeling_fermi_2014}%
  \BibitemOpen
  \bibfield  {author} {\bibinfo {author} {\bibfnamefont {J.}~\bibnamefont
  {Keeling}}, \bibinfo {author} {\bibfnamefont {M.~J.}\ \bibnamefont
  {Bhaseen}}, \ and\ \bibinfo {author} {\bibfnamefont {B.~D.}\ \bibnamefont
  {Simons}},\ }\bibfield  {title} {\enquote {\bibinfo {title} {Fermionic
  superradiance in a transversely pumped optical cavity},}\ }\href {\doibase
  10.1103/PhysRevLett.112.143002} {\bibfield  {journal} {\bibinfo  {journal}
  {Phys. Rev. Lett.}\ }\textbf {\bibinfo {volume} {112}},\ \bibinfo {pages}
  {143002} (\bibinfo {year} {2014})}\BibitemShut {NoStop}%
\bibitem [{\citenamefont {Piazza}\ and\ \citenamefont
  {Strack}(2014)}]{piazza_fermi_2014}%
  \BibitemOpen
  \bibfield  {author} {\bibinfo {author} {\bibfnamefont {Francesco}\
  \bibnamefont {Piazza}}\ and\ \bibinfo {author} {\bibfnamefont {Philipp}\
  \bibnamefont {Strack}},\ }\bibfield  {title} {\enquote {\bibinfo {title}
  {Umklapp superradiance with a collisionless quantum degenerate fermi gas},}\
  }\href {\doibase 10.1103/PhysRevLett.112.143003} {\bibfield  {journal}
  {\bibinfo  {journal} {Phys. Rev. Lett.}\ }\textbf {\bibinfo {volume} {112}},\
  \bibinfo {pages} {143003} (\bibinfo {year} {2014})}\BibitemShut {NoStop}%
\bibitem [{\citenamefont {Chen}\ \emph {et~al.}(2014)\citenamefont {Chen},
  \citenamefont {Yu},\ and\ \citenamefont {Zhai}}]{zhai_fermi_2014}%
  \BibitemOpen
  \bibfield  {author} {\bibinfo {author} {\bibfnamefont {Yu}~\bibnamefont
  {Chen}}, \bibinfo {author} {\bibfnamefont {Zhenhua}\ \bibnamefont {Yu}}, \
  and\ \bibinfo {author} {\bibfnamefont {Hui}\ \bibnamefont {Zhai}},\
  }\bibfield  {title} {\enquote {\bibinfo {title} {Superradiance of degenerate
  fermi gases in a cavity},}\ }\href {\doibase 10.1103/PhysRevLett.112.143004}
  {\bibfield  {journal} {\bibinfo  {journal} {Phys. Rev. Lett.}\ }\textbf
  {\bibinfo {volume} {112}},\ \bibinfo {pages} {143004} (\bibinfo {year}
  {2014})}\BibitemShut {NoStop}%
\bibitem [{\citenamefont {Sandner}\ \emph {et~al.}(2015)\citenamefont
  {Sandner}, \citenamefont {Niedenzu}, \citenamefont {Piazza},\ and\
  \citenamefont {Ritsch}}]{sandner_2015}%
  \BibitemOpen
  \bibfield  {author} {\bibinfo {author} {\bibfnamefont {R.~M.}\ \bibnamefont
  {Sandner}}, \bibinfo {author} {\bibfnamefont {W.}~\bibnamefont {Niedenzu}},
  \bibinfo {author} {\bibfnamefont {F.}~\bibnamefont {Piazza}}, \ and\ \bibinfo
  {author} {\bibfnamefont {H.}~\bibnamefont {Ritsch}},\ }\bibfield  {title}
  {\enquote {\bibinfo {title} {Self-ordered stationary states of driven quantum
  degenerate gases in optical resonators},}\ }\href
  {http://stacks.iop.org/0295-5075/111/i=5/a=53001} {\bibfield  {journal}
  {\bibinfo  {journal} {EPL (Europhysics Letters)}\ }\textbf {\bibinfo {volume}
  {111}},\ \bibinfo {pages} {53001} (\bibinfo {year} {2015})}\BibitemShut
  {NoStop}%
\bibitem [{\citenamefont {Christensen}\ \emph {et~al.}(2008)\citenamefont
  {Christensen}, \citenamefont {Will}, \citenamefont {Saba}, \citenamefont
  {Jo}, \citenamefont {Shin}, \citenamefont {Ketterle},\ and\ \citenamefont
  {Pritchard}}]{christensen2008trapping}%
  \BibitemOpen
  \bibfield  {author} {\bibinfo {author} {\bibfnamefont {Caleb~A}\ \bibnamefont
  {Christensen}}, \bibinfo {author} {\bibfnamefont {Sebastian}\ \bibnamefont
  {Will}}, \bibinfo {author} {\bibfnamefont {Michele}\ \bibnamefont {Saba}},
  \bibinfo {author} {\bibfnamefont {Gyu-Boong}\ \bibnamefont {Jo}}, \bibinfo
  {author} {\bibfnamefont {Yong-Il}\ \bibnamefont {Shin}}, \bibinfo {author}
  {\bibfnamefont {Wolfgang}\ \bibnamefont {Ketterle}}, \ and\ \bibinfo {author}
  {\bibfnamefont {David}\ \bibnamefont {Pritchard}},\ }\bibfield  {title}
  {\enquote {\bibinfo {title} {Trapping of ultracold atoms in a hollow-core
  photonic crystal fiber},}\ }\href@noop {} {\bibfield  {journal} {\bibinfo
  {journal} {Phys. Rev. A}\ }\textbf {\bibinfo {volume} {78}},\ \bibinfo
  {pages} {033429} (\bibinfo {year} {2008})}\BibitemShut {NoStop}%
\end{thebibliography}
\end{document}